\documentclass[preprint]{elsarticle}
\usepackage{lineno,hyperref}
\modulolinenumbers[5]
\journal{Journal of Vacuum}
\usepackage{graphicx}
\usepackage{subcaption}
\usepackage{placeins}
\usepackage{multirow}
\usepackage{color}
\usepackage{amsmath}
\usepackage[utf8]{inputenc}
\usepackage[T1]{fontenc}
\usepackage{mathptmx}
\usepackage{etoolbox}

\begin{document}

\begin{frontmatter}

\title{Experimental investigation of free jets through supersonic nozzles}

\author[affiliation1,affiliation2]{Milaan Patel\corref{correspondingauthor}}
\cortext[correspondingauthor]{Email:milaan.patel@gmail.com}

\author[affiliation1,affiliation2]{Jinto Thomas}

\author[affiliation1,affiliation2]{Hem Chandra Joshi}

\address[affiliation1]{Institute for Plasma Research, Near Bhat, Gandhinagar 382428, Gujarat, India}
\address[affiliation2]{Homi Bhabha National Institute, Training School Complex, Anushaktinagar, Mumbai 400094, India}

\begin{abstract}

In this paper, we report experimental investigation to improve the shape of a supersonic nozzle for rarefied flows to generate high axial density at extended distances from the nozzle.
The reported work is significant for molecular jet/beam applications that require high center-line density and narrow jet profile. 
We investigate a parabolic nozzle whose profile is generated using the virtual source model of free expansion and compare its performance with a set of conical nozzles having different cone angles using simulations as well as experiments. 
All nozzles are made by additive manufacturing using ABS and performance is found to be satisfactory. 
Axial density and lateral spread of the jets are measured using a pitot tube assembly. The accuracy and operational limit of the pitot tube for rarefied flow is quantified by using established mathematical and empirical models for a sonic nozzle. 
The study demonstrates that the performance of the parabolic nozzle is comparable or slightly better as compared to conical counterparts. Moreover, the parabolic profile can be used for optimizing the opening angle for conical nozzles of various lengths.

\end{abstract}


\end{frontmatter}


\section{Introduction}\label{sec:int}

Rarefied free supersonic jets inside a vacuum vessel have applications in many fields \cite{2018EAST,NSTX2019,2016_cluster}. 
High axial density with low divergence is essential for applications like tokamak refueling for better efficiency and in cluster experiments to produce large sized clusters. Experiments on tokamak edge plasma diagnostics \cite{Wendler_2022,kru2012,4_TJII_Atomic_beam} and ion-beam profile monitoring \cite{LHC2022} use supersonic molecular beam extracted from the jet using axially placed skimmer. Here higher beam density is required to improve signal to noise ratio. For applications where molecular beam is used to determine reaction cross sections and gas-surface interactions \cite{2016_gsi,gsi_book}, high directionality is critical. This is achieved with long nozzle-skimmer distances. However, at longer distances, molecules get reflected from the vessel surface and enter inside the jet resulting in the interaction with the jet and thermalizing it \cite{Mil2021}. Hence, it is crucial for the jet to have large axial density at long skimming distances.

Conventionally, axial density is increased by increasing the reservoir pressure. However, this requires higher pumping rate to prevent accumulation of the background gas in the vessel and hence the jet is operated in short bursts. Burst duration has to be limited to the extent when the rising background pressure starts adversely affecting the jet. This imposes a limit that the experiments should be conducted within the available burst duration. Using supersonic nozzle one can get higher axial density to compensate for large flow rate. However, the manufacturing limitations of small sized nozzle often require to adopt simple conical geometry. 

In a conical nozzle, the exit cone angle of the expanding section dictates the maximum centerline density that can be achieved for a particular flow rate \cite{kluria2011}. However, work was limited to short aspect ratio nozzles ($d_{exit} \approx l_{nozzle}$) for which it was found that the optimum half cone angles for high centerline density should be 40-50 deg. On the other hand, in the experiments which use de Laval nozzle, the profile of the expansion section is generated by solving equations of mass, moment and energy using method of characteristics. Generated profile is then scaled down to sub mm size for a supersonic nozzle \cite{NSTX2019,2_lorenz_laser}. This approach has been in use to generate optimum nozzle profile to achieve uniform flow at the nozzle exit\cite{br_1985}. A detail discussion on the estimation of various parameters is given in reference \cite{Durif_2022}. The flow stays uniform beyond the exit provided the exit pressure of the jet ($P_e$) and the background pressure ($P_b$) are equal. However, for rarefied jets, background pressure is atleast an order of magnitude lower and the jet is always “under-expanded” inside the nozzle. Hence, uniform flow cannot be achieved after the exit and one cannot guarantee optimum performance at long axial distances. This indicates that a different nozzle design approach has to be attempted to get higher axial density at longer distances for rarefied jets.

 In this work we consider a supersonic nozzle in which the jet becomes rarefied before it exit the nozzle. Then we try to generate nozzle shape to direct the rarefied jet along the axis and thereby increasing the axial density. We call this arrangement molecular nozzle. Concurrently we aim to achieve narrow density spread along the lateral direction which is important in minimizing the reflections from the walls of the chamber at longer extraction distances of the skimmer subsequently reducing the background penetration. The nozzle profile we use for molecular nozzle is a parabolic profile which is generated using virtual source model of free expansion. We compare its performance with a set of conical nozzles having different opening angles but having same length and flow rate as that of the parabolic nozzle. Using simulations as well as experiments we compare the axial and lateral density profiles of the free expanding flows exiting the nozzles.

Simulations are carried out using Direct Simulation Monte Carlo (DSMC) code by G.Bird \cite{Bird2013}. Experimental measurement of the jet density is carried out by measuring the impact pressure using a pitot tube. The pressure of the jet measured by pitot tube lies mostly in the transient and molecular region. Hence, accurate values of absolute densities could not be estimated using continuum flow assumptions typically used in this approach. On the other hand, molecular flow based methods are applicable only for simple geometries \cite{abm1988} and accurate implementation to pitot tube used in this experiment is not feasible. Additional limitations arise because the measurements are carried out at the limits of sensitivity and resolution of the instruments. As a results, nozzle performance could be only evaluated qualitatively. The trend, of course, is in agreement with DSMC simulations. We have measured the limit of operation of pitot tube using a sonic nozzle and quantified the errors. Molecular nozzles are manufactured using additive manufacturing technique using acrylonitrile butadiene styrene (ABS). The assumption of molecular flow used in designing the nozzle requires the internal surface of the expanding section to be smooth. This makes controlled post processing of the nozzle an important factor. These aspects along with vacuum performance of the nozzle material are discussed in the work.

The paper is organized in following way: Theory and design approach are discussed in section 2, experiment setup is described in section 3, simulation conditions are discussed in section 4, results are discussed in section 5 and section 6 concludes the work.

\section{Theory and design approach}\label{sec:the}

In this section we briefly discus continuum theory of supersonic expansion of free jet from a sonic nozzle. We then compare it with semi-empirical angular density distribution model for rarefied jet \cite{bei1981} which is based on virtual source assumption \cite{ash1966,she1963}. We use underline assumptions of the virtual source model to generate parabolic nozzle. However, we limit our discussion to light monatomic gas helium because of additional complexity involved with heavier monoatomic, diatomic and polyatomic gases.

\subsection{Free supersonic expansion}

Supersonic source used in this experiment consists of high pressure gas expanding isentropically inside a vacuum vessel through a sonic nozzle. This jet always behaves as a highly under-expanded supersonic jet with free expanding supersonic core inside the shock boundary. One can decrease the background pressure ($P_b$) of the vessel enough so that continuum flow does not prevail at jet boundaries and the shock structure is absent. In such a case, continuum flow is limited to a certain distance after the nozzle exit. The streamlines of the flow after exiting the nozzle diverge and become straight after a certain distance known as freeze plane \cite{abm1988}. The flow, then, becomes molecular and the subsequent expansion is purely geometric. Molecular beams are generated by placing the skimmer farther from the freeze plane where the flow is molecular. Properties of the extracted supersonic molecular beam from an ideal skimmer can be derived from the properties of the parent jet using nozzle-skimmer distance and geometries of nozzle and skimmer. 

In the continuum region of expansion, one can numerically solve differential equations for momentum and energy along with the equations of state by assuming isentropic expansion to get Mach numbers at different axial and angular positions from the nozzle. All the remaining jet parameters e.g. density, temperature and pressure can be calculated assuming $1D$ expansion along the streamlines. Same approach can be adopted for a supersonic nozzle as well. However, flow through supersonic nozzle is accompanied by viscous boundary layer in the diverging part of the nozzle and it is rather difficult to take account of the viscous effects. On the other hand, viscous effects are absent in sonic nozzle as there is no diverging part and hence the above approach can give realistic results. The axial Mach number in terms of nozzle diameter for free supersonic expansion through a sonic nozzle in continuum region is given by equation \ref{eq:mach} \cite{abm1988}. 

\begin{eqnarray}
M=A\left(\frac{z-z_0}{d}\right)^{\gamma-1} - \frac{\frac{1}{2}\frac{\gamma+1}{\gamma-1}}{A\left(\frac{z-z_0}{d}\right)^{\gamma-1}} 
\label{eq:mach}
\end{eqnarray}

Here, $\gamma$ is the ratio of specific heat of gas at constant pressure and constant volume, $z/d$ is the axial distance from the throat normalized to the throat diameter, $z_0/d$ is the point where straight streamlines appear to radiating out from and its value is specific to gas type (0.075 for monoatomic and 0.4 for diatomic). $A$ is constant which also depends on gas type (3.26 for monoatonic and 3.65 for diatomic).
The viscous effects in the converging section of sonic nozzle can be accounted by using coefficient of discharge ($C_d$) which is the ratio of the experimental to theoretical flow rates. For sharp sonic nozzle, one can assume $C_d\approx90\%$. In our study we take into account viscous effects by measuring the flow rate ($\dot{N}$) experimentally and calculating the reduced nozzle diameter ($d$) using eq. \ref{eq:flowrate}

\begin{eqnarray}
\dot{N}=f(\gamma)n_0\alpha_0\pi (d/2)^2
\label{eq:flowrate}
\end{eqnarray}

Where, $n_0$ is reservoir number density, $\alpha_0$ is the most probable thermal speed of the molecules at reservoir temperature. Detailed description is given in (\ref{appA}). Once the Mach number is known, rest of the jet parameters can be calculated using a set of 1D isentropic expansion equations along the streamlines (equations-\ref{eq:isentriopic}). Here, the subscript '0' represents reservoir conditions, $n$ is the number density, $T$ is the temperature, $u$ is the velocity and $u_{\infty}$ is the terminal velocity.

\begin{subequations}
\begin{gather}
\frac{T}{T_0}=\left(1+\frac{\gamma-1}{2}M^2\right)^{-1} \label{eq:t} \\
\frac{n}{n_0}=\left(1+\frac{\gamma-1}{2}M^2\right)^{\frac{-1}{\gamma-1}} \label{eq:n} \\
\frac{u}{u_{\infty}}=\left(1-\frac{T}{T_0}\right)^{1/2} \label{eq:v}
\end{gather}
\label{eq:isentriopic}
\end{subequations}

At longer distances from the nozzle, where molecular flow is achieved, the streamlines become straight and appear to emit radially outward from a single point which can be considered as a virtual source \cite{ash1966,she1963}. Semi-empirical relation given by Beijerinck et al. \cite{bei1981} can be used to estimate axial-angular number density (eq.-\ref{eq:n_bjrnk}).
 
\begin{eqnarray}
n(z,\theta)=\frac{\kappa \dot{N}}{u_{\infty}\pi z^2} cos^b\left(\frac{\pi}{2}\frac{\theta}{\theta_{PM}}\right)
\label{eq:n_bjrnk}
\end{eqnarray}

\begin{eqnarray}
\theta_{PM}=\frac{\pi}{2} \left[\left(\frac{\gamma+1}{\gamma-1}\right)^{1/2}-1\right]
\label{eq:tpm}
\end{eqnarray}

Where, $z$ is the axial distance, $\dot{N}$ is the flow rate, $\theta$ is the angle with respect to the flow axis and $\kappa$ is the "peaking factor" which represents the effectiveness of the nozzle when compared to the effusive source. ($\kappa$=1 for effusive source). Constant $b$ represents the angular distribution of density. Higher values of $b$ indicate higher fall rate of the density at larger angles with respect to the axis. Again the values of $\kappa$ and $b$ are gas dependent and are given in table-\ref{table:ndconstants}.

\begin{table}[h]
\begin{center}
\begin{tabular}{|c|c|c|}
\hline
$\gamma$& $\kappa$ 	& $b$	\\ \hline
5/3 	& 1.98 		& 3 	\\ \hline  
7/5		& 1.38 		& 4.32 	\\ \hline
9/7		& 1.10 		& 5.47  \\ \hline  
\end{tabular}
\end{center}
\caption{\label{table:ndconstants} Peaking factor and angle factor for monoatomic, diatomic and polyatomic gas \cite{bei1981}.}
\end{table}

Further, viscous effects can be accounted for by measuring the flow rate experimentally. We measure the flow rate using pressure rise method (see \ref{appB}). For the present experiment it is found to be $(80\pm5)\%$ of the theoretical value. It is slightly smaller than the typical coefficient of discharge for sharp sonic nozzle ($C_d\approx 90\%$) as orifice used in the experiment is not sharp ($l \approx 2d$).
Briefly, eq.-\eqref{eq:mach} is used for continuum jets whereas eq.-\eqref{eq:n_bjrnk} represents empirical model for rarefied jets.  
   
\subsection{Parabolic profile of molecular nozzle}

Flow velocity reaches $98\%$ of the terminal velocity within first few nozzle diameters ($x/d \approx 5$, from equations \eqref{eq:mach} and \eqref{eq:isentriopic}). For rarefied jets, this is also typically the region where transition from continuum to molecular flow occurs \cite{abm1988}. As a result, the streamlines of the molecular flow become straight for $x/d>5$ and continue to travel in the direction in which they are emitted. However, by introducing a smooth solid surface in its path, it is possible to change the direction. By contouring the surface appropriately, we can direct the expanding streamlines in the desired direction. Ideally, after reflection, streamlines should follow the new direction due to low collision frequency.

This leads us to two important assumptions to ensure the working of such a contoured surface. (1) The collision frequency of the gas molecules should be sufficiently small to avoid any significant change in the path of the streamline before and after reflecting from the surface. (2) the gas surface collision should be elastic. Condition of first assumption is relatively easy to achieve by decreasing the reservoir pressure sufficiently so that molecular flow is achieved after a few exit diameters. However, care should be taken that pressure is high enough that the flow does not become effusive. The second assumption cannot be achieved in a realistic scenario but one can come very close by choosing combination of lighter gases and heavy surface molecules. For rarefied flows, gas-surface interaction is represented by momentum and energy accommodation coefficients \cite{Bird1994}. Both coefficients scale from 0 to 1. Higher values indicate diffused reflection, whereas lower values specular reflection. For supersonic flow of lighter gas over the heavier surface, values of these coefficients are smaller. For instance, in case of helium interacting with graphite, momentum and energy accommodation coefficients are typically below 0.15 and 0.30 respectively \cite{2019_accomodation}. Hence, helium gas is more prone to reflect elastically than other heavier monoatomic and diatomic gases. Briefly, contoured surface concept works better with helium as compared to other heavier gas. 

If the streamlines of the flow originate from a virtual source, a parabolic reflecting contour surface with focal point coinciding with the virtual source location can be used to focus the streamlines parallel to the axis. Thus, the axisymmetric rotation of the parabolic contour forms a molecular nozzle. The contour surface reflects all the streamlines along the axis except the one which lies within the solid angle subtended by the exit opening of the nozzle. One can increase the envelop of the reflecting surface or decrease the solid angle of undisturbed jet by increasing the nozzle length. Hence, longer nozzles work favorably in directing major part of the flow axially.

\begin{figure}[ht]
\centering
\includegraphics[width=0.8\columnwidth, trim=0cm 0cm 0cm 0cm, clip=true,angle=0]{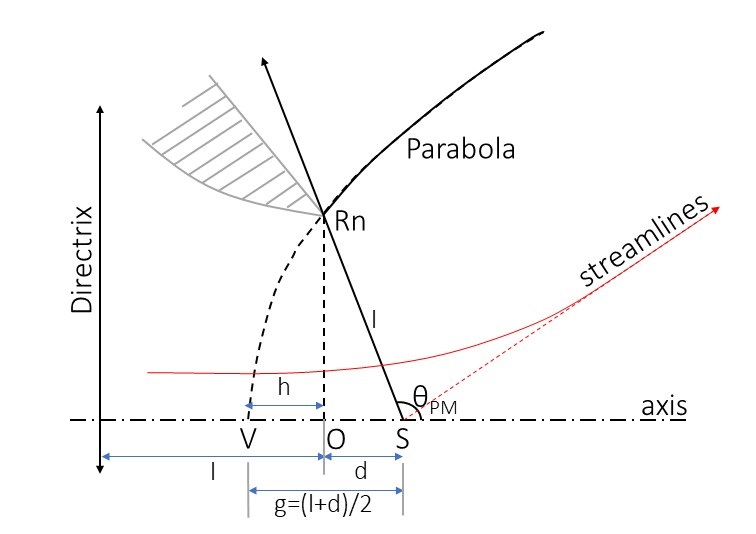}
\caption{\label{fig:parabola} Parabolic profile is defined as the locus of the points which are equi-distance from the directrix and the focus(S). This is used to calculate constants $g$ and $h$ using basic trigonometry. V is the vertex of parabola, O is the origin of the (r,z) coordinate system and $R_n$ is the throat radius.}
\end{figure}

As shown in figure (1), parabola can be described by two parameters i.e. (i) the location of the virtual source that coincides with the focal point of the parabola and (ii) the nozzle radius $R_n$. The location of the virtual source $S$ can be determined from the geometry using Prandtl-Meyer relation (eq. \ref{eq:tpm}). Hence, the generated parabolic profile is specific to the throat radius ($R_n$) and the type of gas ($\gamma$). The parabolic profile can be expressed in $r$, $z$ coordinate using eq. \ref{eq:parabolic}. 

\begin{subequations}
\begin{gather}
r^2=4g(z-h), z>0 \label{eq:parabola} \\
g=\frac{R_n}{2}\left(\frac{1-cos\theta_{PM}}{sin\theta_{PM}}\right)\label{eq:parabolic_g} \\
h=-\frac{R_n}{2}\left(\frac{1+cos\theta_{PM}}{sin\theta_{PM}}\right) \label{eq:parabolic_h} 
\end{gather}
\label{eq:parabolic}
\end{subequations}

Taking origin of the coordinate system at the intersection of the sonic plane and the axis of the nozzle, constants $g$ and $h$ can be determined from the geometry of the parabola.

\section{Experiment setup}\label{sec:exp} 

\begin{figure}[ht]
\centering
\includegraphics[width=1.0\columnwidth, trim=0cm 0cm 0cm 0cm, clip=true,angle=0]{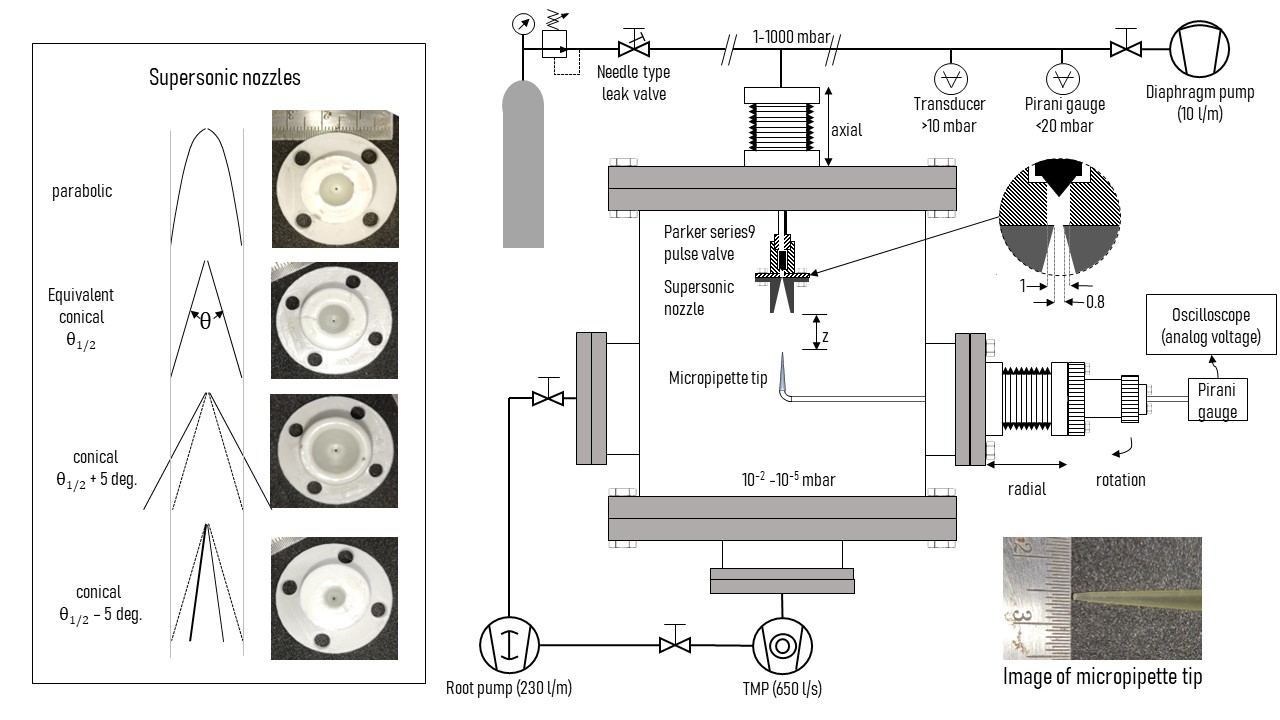}
\caption{\label{fig:exp_setup} A schematic view of the experimental setup used for density measurements.}
\end{figure}

The experimental setup is shown in fig. \ref{fig:exp_setup}. A cylindrical vacuum vessel of diameter 300 mm and 320 mm long and oriented vertically is used as a test vessel to generate the free supersonic jet. The vessel is pumped by Pfeiffer Hi-pace 700 turbo-molecular pump (pumping speed for helium is 650 $l/s$ at $<10^{-3}$ mbar) mounted at the bottom (DN 160 CF flange) of the vessel. A dual stage root pump Pfeiffer ACP-15 (with pumping speed 230 $l/m$ at 1 mbar) is used as backing pump to TMP and for rough vacuum generation ($>0.1 mbar$) inside the vessel. Pressure inside the vessel is monitored using Pfieffer PKR251 combined full range (Pirani + cold cathode ionization gauge) with an accuracy of 30\%. 

A Parker series 9 pulse valve with 1 mm orifice is used as a source of gas feed. It is mounted inside the vessel using 1/4 inch SS-tube via UHV gas feed-through. The feed-through is mounted on a bellow type manually controlled UHV translation stage to move the valve axially. Length scale of the translation stage has least count of 0.1 mm which we can assume as accuracy achieved in axial positioning by eyeballing the markers. Ultra high purity (99.999\%) helium gas is supplied to the valve at pressure ranging from 1-1000 mbar through a low pressure gas supply line. The gas supply line is pumped via self-sealing Pfieffer made diaphragm vacuum pump (10 $l/m$) to prevent contamination from the backflow. Gas supply from high pressure helium cylinder bottle is regulated using a 	precision needle valve connected immediately after the gauge pressure regulator. The pressure inside the gas supply line is monitored by a diaphragm type DMP 320 pressure transducer with 1\% accuracy for pressure above 10mbar and Pirani gauge with 10\% accuracy for pressure below 20mbar (calibrated for helium gas).
 
Impact pressure is measured using a pitot tube assembly. It consists of a micropipette tube of 0.5 mm orifice and <200 micron edge sharpness hot sealed on 1/4 inch SS tube which extends outside the vacuum vessel via UHV gas feed-through. The outside end of the SS tube is connected to a Pirani gauge (MKS instruments micropirani$^{TM}$ , accuracy 10\%). The gas feed-through is mounted on another bellow type UHV translation stage. Additional rotational coupling is provided for in-situ alignment with the pulse valve. The SS tube, micro-pipette tip, gas feed-trough, translation stage and Pirani gauge forms the pitot-tube assembly. Analog voltage signal from the Pirani gauge is recorded using an oscilloscope. This is done to achieve fast readout time to ensure dynamic stability of pressure which is not possible from the digital display panel of the gauge controller. Voltage signal is converted to pressure using calibration provided by the manufacturer. 

Pulse valve is operated (by IOTA-One controller) in continuous open mode at nozzle pressure <20mbar and in semi-continuous mode with duty cycle of 15-20 s for the pressure range 20-100 mbar to limit the gas load on the Turbo molecular pump. In semi-continuous mode duty cycle is always higher than stabilization time of the pitot tube assembly and the gas supply line. Approximately, 10 s are required to stabilize the pitot tube assembly in molecular flow region (<2 s for continuum flow) and 10 s to stabilize the pressure of gas supply line. 

\subsection{Additive manufacturing of parabolic nozzle}

Supersonic nozzles used in the experiment are mounted on the pulse valve of 1 mm orifice. All the nozzles have 0.8 mm throat diameter and expansion section is 20 mm long. Orifice diameter of the nozzle is kept smaller than the pulse valve to ensure sonic conditions at the nozzle. The nozzle is additively manufactured by fused deposition method (FDM) using acrylonitrile butadiene styrene (ABS). In FDM, the thermoplastic material is fed through a heated moving head that extrudes it and deposits it layer by layer in the desired shape. Layer thickness of the nozzle is kept 0.1 mm to considering the initial surface finish and strength. Larger layer thickness results in better strength but in lower surface finish and vise-versa. ABS is slightly difficult to work with FDM machines due to its thermal contraction. However, it is selected for its ease in post processing. The factors considered while selecting a thermoplastic material ABS are thermal stability, structural strength, surface finish, out-gassing, and surface erosion. 

Thermal stability of the ABS is not a concern for room temperature jets as temperature inside the nozzle is always less or equal to the stagnation temperature. (Heat deflection temperature of ABS is $\approx 70^oC$). At room temperature, ABS is also one of the strongest thermoplastic materials. For nozzle pressure below 1000 mbar, the differential pressure is quite low to have any kind of structural effects. However, surface finish is the major concern in the nozzle fabricated using the FDM process. FDM fabricated surfaces have visible layer lines. To remove these layer lines and make surface smooth we use acetone vapor smoothening. As ABS is soluble in acetone, when the surface of the nozzle is subjected to acetone vapor, it partially dissolves the surface of the nozzle in contact. The dissolved slurry levels the discontinuities between the layers and evens out the surface. However, disadvantage of this method is that over the time acetone is absorbed into the bulk material making it soft like a jelly. This deforms the structure under its own weight. To prevent softening, vapour smoothing was performed in short cycles. In single cycle the surface was subjected to acetone for 1 minute by suspending the parts approximatly 2 cm from the pool of acetone maintained at $(40^oC)$. In the next step, acetone is evaporated out from the component by placing it in vacuum desiccator (1 mbar) for 1 hour. The cycle is repeated until smooth (glass like) surface finish is achieved. The number of cycles vary largely depending on the geometry. In our case optimum number of cycles is found to be 10. Further, the surface finish remains consistent during and after the experiments and no deterioration of the surface finish or erosion was observed. 

One of the detrimental factors associated with the utilization of plastic parts in high vacuum is their relatively high outgassing and degassing rates compared to metals. The degassing rate can be minimized using a long pumping time which in our case is approximately 24 hours. However, plastic material outgass for long periods due to steady boiling of the plastizers\cite{jen1956}. These can be quantified in terms of vapor pressure. For polymers at room temperature, vapor pressure typically lies below $10^{-5}$ mbar. This is an order of magnitude lower than the minimum pressure inside the nozzle during experiment. Thus, the effect of degassing and out-gassing can be neglected in our measurements.

\subsection{Experimental measurement of density}

Experimental value of Mach numbers on the flow axis are measured using pitot tube. In a supersonic continuum flow, a normal shock is formed in front of the pitot tube and it measures the impact pressure of the subsonic flow downstream of the normal shock. Assuming 1D isentropic flow along streamlines (valid on the jet axis), impact pressure can be used to calculate Mach number of the supersonic flow upstream of the normal shock using Rayleigh pitot (RPT) formula as given in eq. \ref{eq:rpt}, where $P_i$ is the impact pressure and $P_{0}$ is the stagnation pressure of supersonic jet which can be assumed same as the reservoir pressure \cite{Durif_2022}.

\begin{eqnarray}
\frac{P_i}{P_0}=\left(\frac{(\gamma+1)M^2}{(\gamma-1)M^2+2}\right)^\frac{\gamma}{\gamma-1} 
\left(\frac{\gamma+1}{2\gamma M^2-\gamma+1}\right)^\frac{1}{\gamma-1}
\label{eq:rpt}
\end{eqnarray}

All the remaining flow parameters on the axis can be calculated from Mach number using the isentropic flow relations. However, eq. \ref{eq:rpt}  cannot hold for molecular flows, because, the formation of aerodynamic shock requires sufficient collisions to define flow parameters like pressure and temperature within the length scale of the apparatus\cite{abm1988}. For molecular flows, the aerodynamic shocks behave as the region of large number of collisions due to scattering from the nearby surface. This creates interference effect similar to one observed in skimmers \cite{shockwave2017} which reduces the throughput of gas through pitot tube. Hence, the impact pressure measured in molecular flow would be the stagnation pressure of the supersonic part of the jet minus pressure loss due to interference at the scattering region on the tip.

We would like to mention that in the present work, RPT equation is used only to validate experiment setup using a sonic nozzle of 1 mm diameter. The reservoir pressure is varied from 100-5 mbar. The upper limit of the reservoir pressure is set considering gas load on the Turbo molecular pump, where as, the lower limit of the reservoir pressure is set to 5 mbar because, the accuracy of the Pirani gauge used to measure impact pressure as well as the conductance of the pitot tube assembly starts deteriorating for pressure below 0.05 mbar (For reservoir pressure lower than 5 mbar the jet pressure reduces below 0.05 mbar making it impossible to measure). At higher reservoir pressures where the jet is in continuum, axial Mach numbers of the expansion are estimated and compared with the prediction of eq.-\eqref{eq:mach} to validate the experiment setup. Reservoir pressure is then reduced in steps and axial pressure is measured for each step. Pressure for which the experimental value deviates from theory is considered as transition from continuum expansion.

For the supersonic nozzle used in this work, the reservoir pressure is maintained below the transition pressure to ensure that the expansion inside the nozzle is in the transient and free molecular region. As the RPT equation can not give exact parameters, the performance of different nozzle profiles are evaluated qualitatively from impact pressure. 

\section{Simulation domain and boundry conditions}\label{sec:sim}

In this section, we give a brief description of the boundary conditions and approximations used in DSMC simulations. In addition, Computational Fluid Dynamics (CFD) simulations using ANSYS are also carried out for the special case of free jet from a sonic nozzle. This is done to compare the flow in continuum and rarefied regimes with the mathematical and semi-empirical models. CFD simulations are not done for supersonic nozzle as those nozzles involve rarefied flow. 

DSMC simulations are carried out by 2 dimentional DS2V code of G. Bird which is primarily used to simulate transient and free molecular flows \cite{Mil2021,kluria2011}. DS2V can also provide a good representation of near continuum flow with Knudson number as small as 0.01 provided the surface collisions are limited\cite{Bird1994}. This makes the DSMC approach more appropriate to simulate free jets for flow conditions that encompass both transient and free molecular flow domains. However, the inherent problem associated with DSMC simulation (or any molecular dynamics method) is extensive computation time to reach an acceptable solution. The selection of the initial number of particles has a significant effect on the computation time to reach the final solution. In DSMC simulation, each simulated molecule represents a group of real molecules. Selecting a large number of simulated molecules can improve the accuracy, however, at the cost of long computation time. Hence, it is crucial to define the initial number of particles to be simulated that provides acceptable results in optimum time. For flows involving large pressure ranges, DSMC simulation scales the number of simulated particles across the whole pressure range linearly. Hence, it is necessary to have smaller number of particles for better computational efficiency for high-pressure region while maintaining sufficient number of particles for the low-pressure region. Considering this, we have not modeled the high pressure reservoir, instead we introduce the flow directly at calculated sonic conditions. For current simulations, we start with approximately 1 million particles by limiting the initial memory allocation for solving. This, eventually scales as the simulation progresses and saturates at 2-5 million particles. Simulation is continued until variation in the number of particles in the flow domain is no more than three standard deviation. Typical computation time required for this simulation on an i7-9750H processor is 5 hours.

\begin{figure}[ht]
\centering
\includegraphics[width=1.0\columnwidth, trim=0cm 0cm 0cm 0cm, clip=true,angle=0]{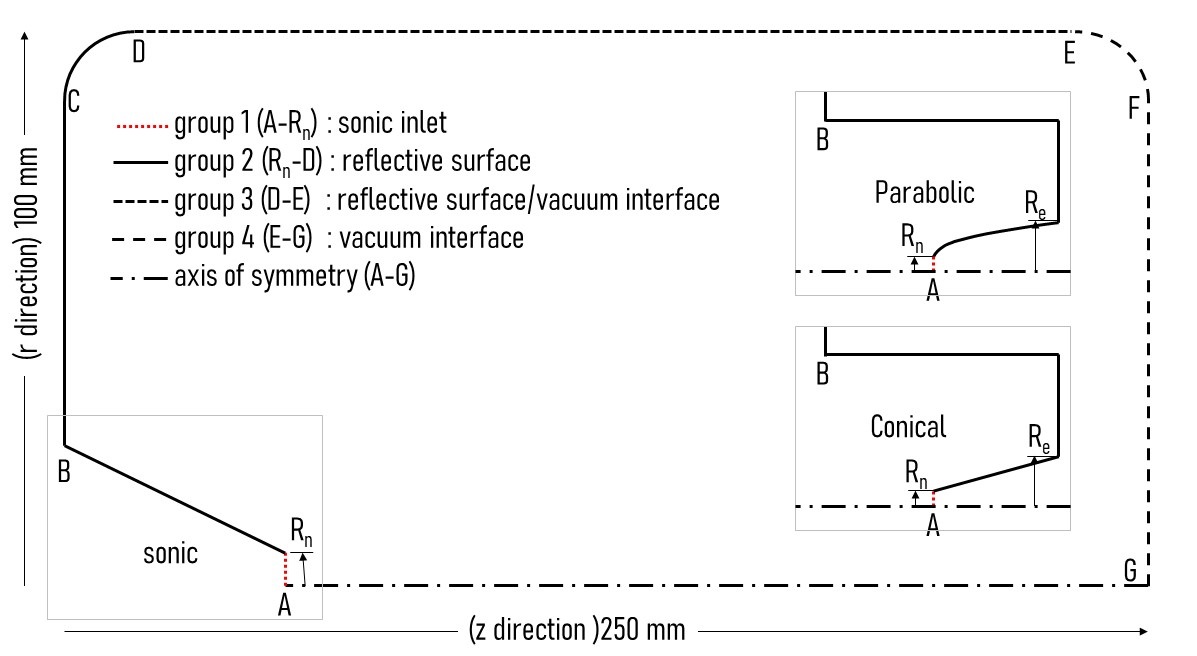}
\caption{\label{fig:fd} Flow domain in simulation is defined by 104 finite intervals(elements) forming surface(A-G) and the axis of symmetry. The intervals forming surface geometry are divided into 4 groups as labeled in the figure. The boundary conditions applied on each group are indicated. Insets are the geometrical shapes for parabolic and conical geometries. $R_n$ is the nozzle radius of the sonic nozzle (throat radius for supersonic nozzle) and $R_e$ is the exit radius of supersonic nozzle.}
\end{figure}

The flow domain used in DSMC simulations is an axisymmetric cross-section of radius 100mm and length 250mm as shown in figure-\ref{fig:fd}. Geometry and flow volume are defined by the bounding surface (A-G) and the axis of symmetry. To simulate flow from a supersonic nozzle, surface between A-B is replaced with appropriate parabolic or conical geometries. The entire bounding surface which forms the geometry is defined by finite intervals (elements described in DSMC Code) which are divided into 4 groups as shown in figure \ref{fig:fd}. The flow boundary conditions are applied on each individual group of intervals. Number of intervals are kept higher at the curvatures and at the regions where large density gradients are expected. Flow inlet boundary condition is applied for the elements of group 1. Flow is introduced with velocity, temperature, and number density corresponding to sonic values calculated from the reservoir conditions assuming isotropic expansion. Intervals of group 2 form the nozzle geometry in case of a supersonic nozzle. Spatially reflective boundary condition is applied for them. This is analogous to an adiabatic surface in which the reflected molecules will have same thermal velocity as that of the incident molecules. This is valid in the case of a rarefied flow for very high Mach numbers over a smooth surface (section 5.8)\cite{Bird1994}. Elements in group 3 form cylindrical boundary surface and elements of group 4 form planar boundary surface at the end. Both boundaries are applied with vacuum interface boundary condition by making them 100\% absorbing.

Simulations for supersonic nozzles are carried out for helium gas at reservoir pressure $P_0$ same as in the experiments ($10 mbar /2.54\times10^{23} molecules \cdot m^{-3}$) expanding through an orifice of diameter 0.8 mm in the background of $10^{-4}$ mbar. For the reservoir pressure of $10$ mbar, the values of pressure, number density, temperature and velocity for sonic conditions are $5$ mbar, $1.59\times10^{23} molecules \cdot m^{-3}$, 224 K and 882 m/s respectively. These conditions are applied at flow inlet boundary.

CFD simulation is carried out for axis-symmetric flow domain  similar to that in figure \ref{fig:fd}. Sonic nozzle is modelled as an orifice with length $\approx 2d$ (as per the schematic diagram of parker series 9 pulse valve). Constant pressure boundary conditions are applied at the inlet and exit. Inlet pressure is taken as the reservoir pressure and exit pressure as background pressure. Walls of the nozzle are considered adiabatic. Viscous effect is applied using realizable $k-\epsilon$ viscosity model. 

\section{Results and Discussion}\label{sec:results} 

In this section, first we discuss the comparison between the theoretical models of continuum and rarefied expansion through a sonic nozzle. We also compare it with CFD simulation and DSMC simulation to see the axial density trends in both the regimes.
We then discuss the validation experiments using sonic nozzle for continuum and rarefied jets. Thus, we assess the validity of pitot tube measurements for different regimes and define the limits of experiment. Considering the limitations, we compare the performance of parabolic nozzle with a set of conical nozzles using DSMC simulation and experiments.

\subsection{Free expansion through a sonic nozzle in continuum and rarefied regime} 

The aim of this study is to see if there is any difference in the behavior of the free jet through a sonic nozzle in continuum and rarefied regimes. We make the comparison between the results of mathematical models and simulations for expansion through a sonic nozzle with orifice ($l=2d$) of 1 mm diameter. 

\begin{figure}[ht]
\centering
\includegraphics[width=0.45\columnwidth, trim=0cm 0cm 0cm 0cm, clip=true,angle=0]{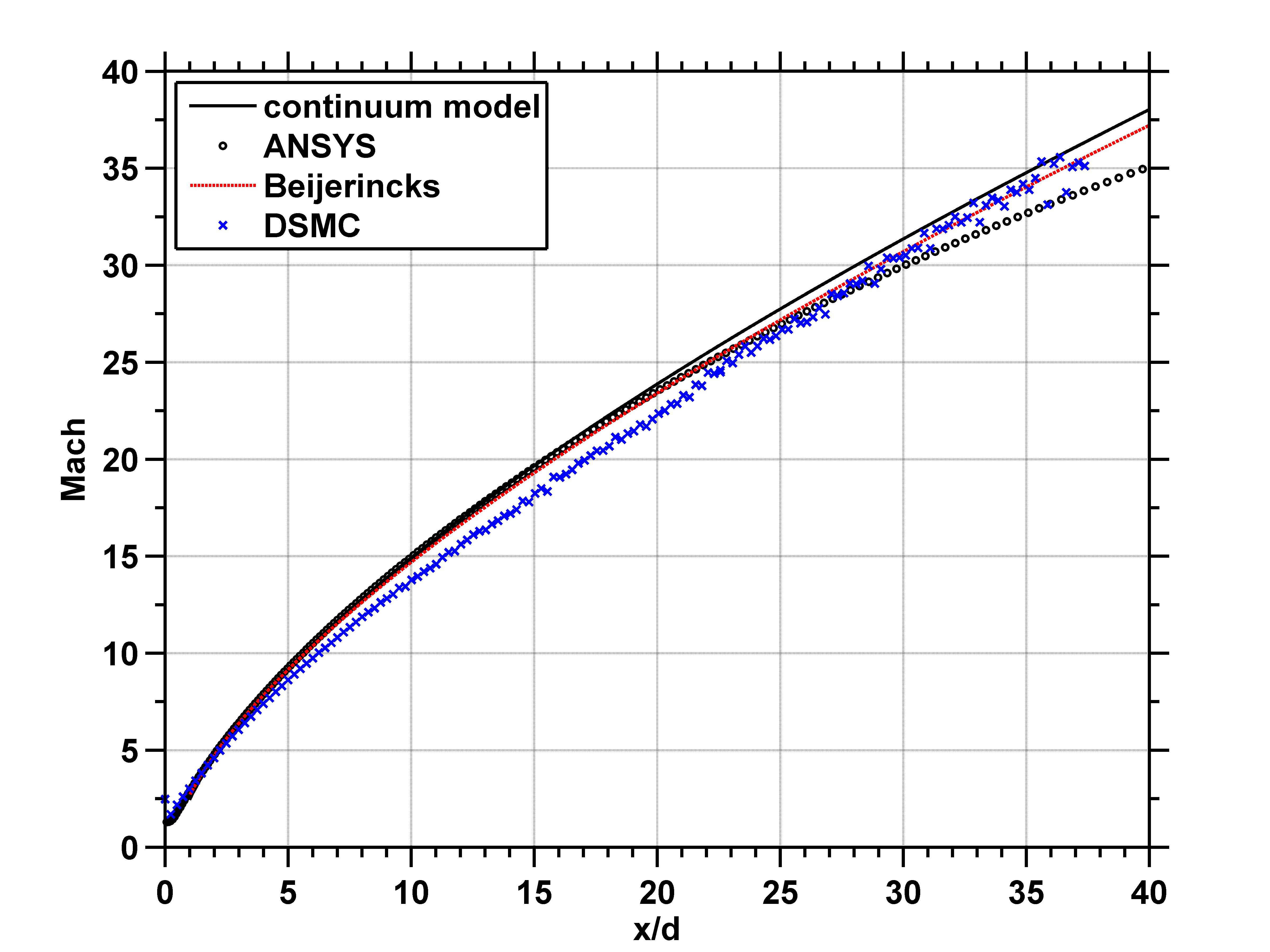}
\includegraphics[width=0.45\columnwidth, trim=0cm 0cm 0cm 0cm, clip=true,angle=0]{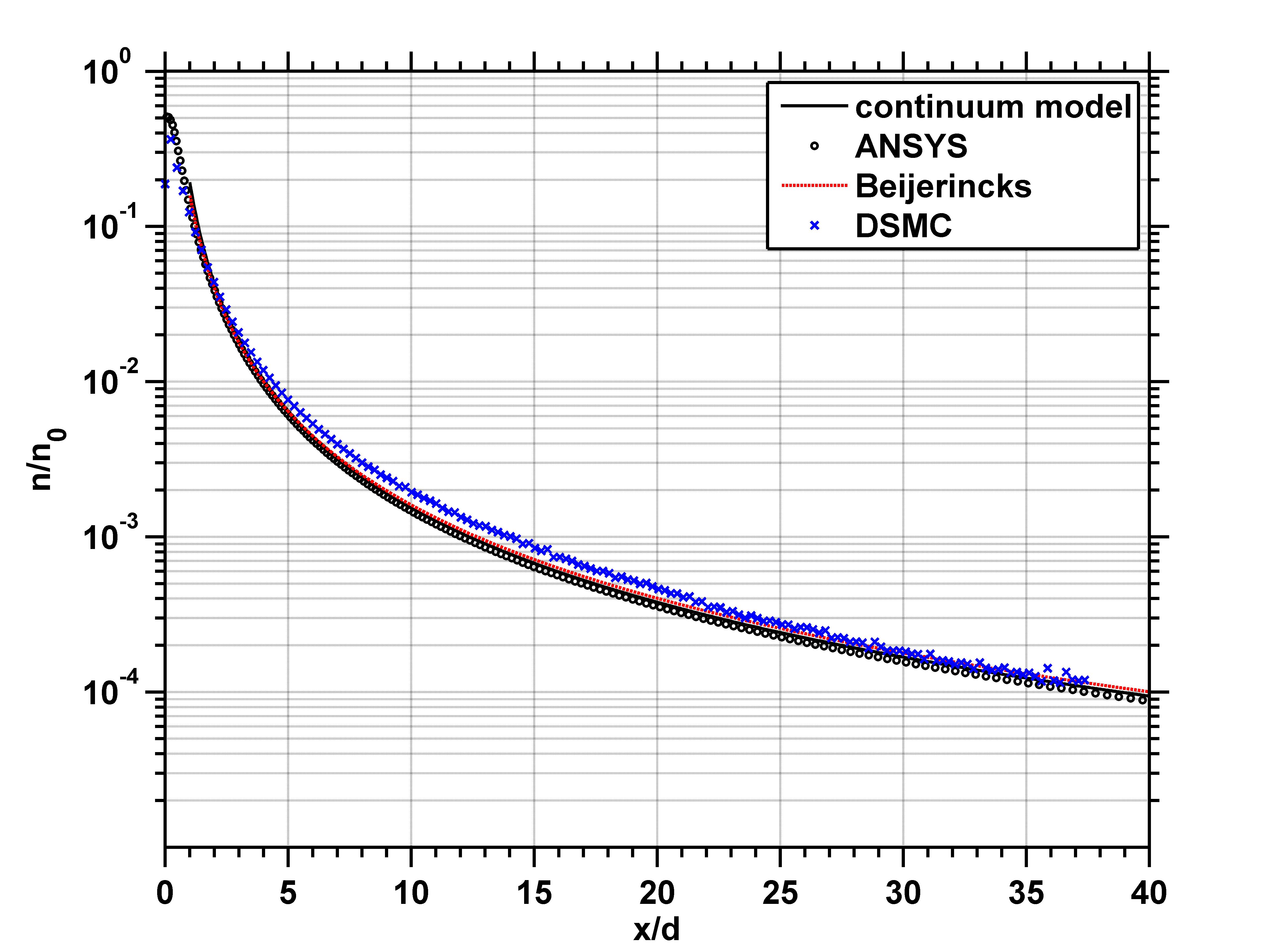}
\caption{\label{fig:cont_rare}Comparison between continuum and rarefied free jets through a sonic nozzle. Plot on left shows axial Mach number and on the right shows number density ($n$) normalized to reservoir number density ($n_0$). Solid lines indicate theoretical predictions. Circular markers shows the values from CFD simulation using ANSYS (for continuum jet) and cross markers values from simulation using DSMC Code(for rarefied jet).}
\end{figure}

For continuum jet, the reservoir pressure and temperature are taken as 50 bar and 300 K, respectively. The values of Mach numbers on flow axis are calculated from eq.-\eqref{eq:mach} and number density from eq.-\eqref{eq:n}. Under the same conditions, CFD-simulation is also carried out as described in section-\ref{sec:sim}. One can calculate the mean free path to ensure that for the length scale of the nozzle diameter, $K_n < 0.02$ upto $40 z/d$ from the nozzle. 
To model the rarefied jet, the reservoir pressure and temperature are taken as 10 mbar and 300 K respectively. Number density is calculated using eq.-\eqref{eq:n_bjrnk} and Mach number is back calculated from the number density using eq.-\eqref{eq:n}. The conditions are simulated using DSMC code as described in section-\ref{sec:sim}. For the rarefied jet, $K_n>1$ after $x/d>5$ , which indicates that the expansion is mostly in the molecular region.  

Figure \ref{fig:cont_rare} shows theoretical and simulated values of axial Mach numbers and densities for both continuum and rarefied jets. Number density ($n$) is normalized to source density ($n_0$) so that different density ranges can be compared. Solid lines represent theoretical values and markers simulated values. One can see that the jet behaves in the same manner for both continuum and rarefied regimes. Hence eq.-\eqref{eq:mach} or eq.-\eqref{eq:n_bjrnk} can be used for the theoretical prediction of number density.

\begin{figure}[ht]
\centering
\includegraphics[width=0.45\columnwidth, trim=0cm 0cm 0cm 0cm, clip=true,angle=0]{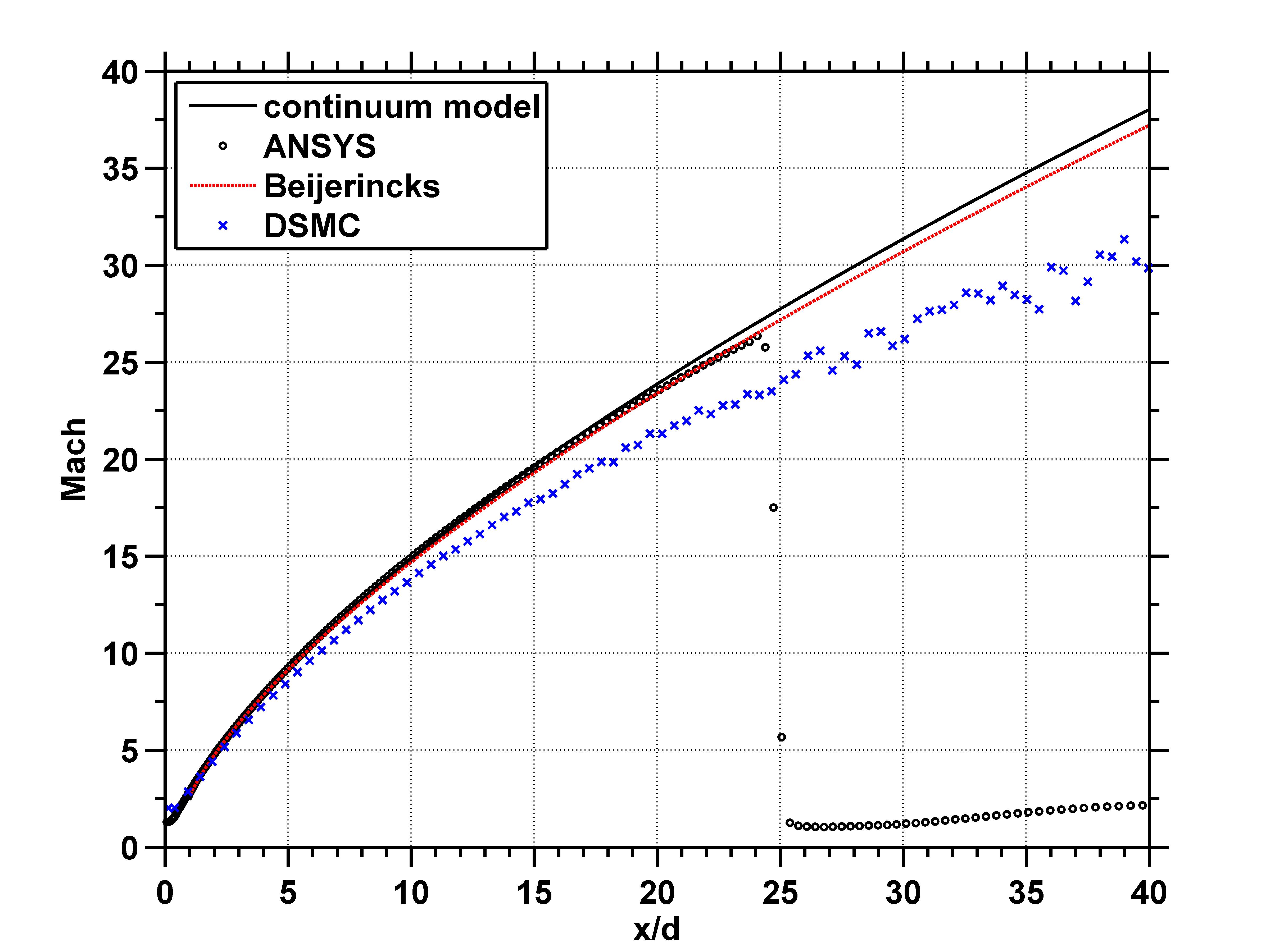}
\includegraphics[width=0.45\columnwidth, trim=0cm 0cm 0cm 0cm, clip=true,angle=0]{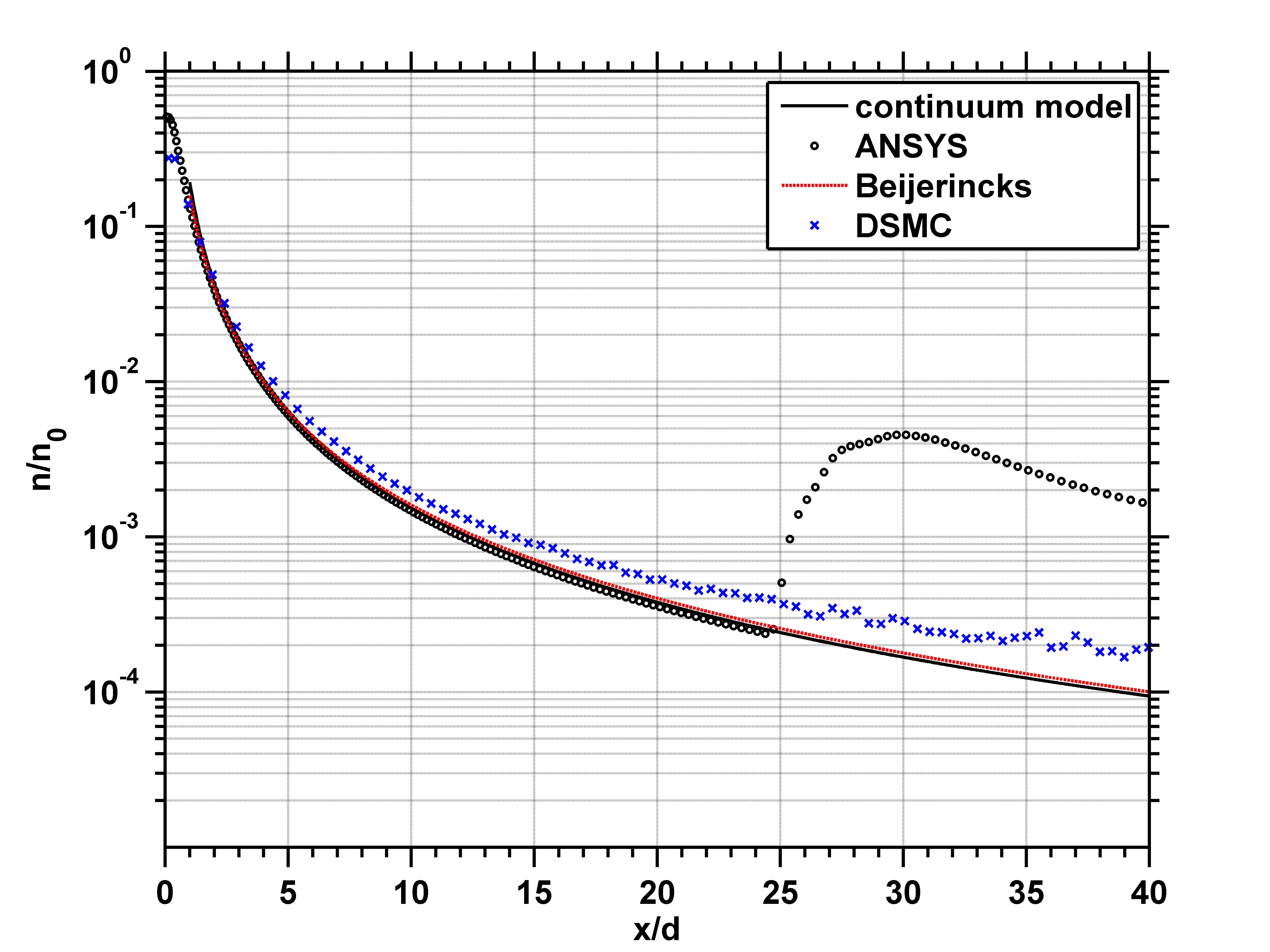}
\caption{\label{fig:cont_rare_sw} Comparison between continuum and rarefied free jets through a sonic nozzle in the presence of background gas.}
\end{figure}

Figure-\ref{fig:cont_rare_sw} shows the jet under same reservoir conditions but in the presence of background gas. The continuum jet (ANSYS simulation) adjusts to the background pressure via shockwave which can be seen as a sharp rise in density (at $x/d\approx 25$). For rarefied jets (DSMC simulation), however, the flow remains supersonic due to the absence of shockwave and number density gradually adjusts to the background density. As mathematical (continuum) and semi-emperical (eq.\eqref{eq:n_bjrnk}) models of free expansion do not account for the background hence it does not show any deviations. In case of continuum jet, the supersonic part is shielded from the background gas by shock wave while in rarefied jet background gas penetrates inside the jet, resulting in partial thermalization. As a result Mach numbers reduces gradually with distance. Similar observations have been noticed in our earlier work \cite{Mil2021} where measured velocity of the rarefied helium jet was found to be lower than expected. 

\subsection{Experimental measurements of axial density for sonic nozzle} 

As RPT equation is valid for continuum flow, the experiments using sonic nozzle were carried out to validate the experimental setup in continuum flow and to determine the extent of deviation to be expected when operating condition of jet changes from continuum to transient and ultimately to rarefied. This helps to determine the lowest limit of the reservoir pressure for reliable measurements of density of the jet with the current setup. Experimentally, the axial density is calculated by measuring the impact pressure and subsequently converting it into Mach number using RPT equation. The measured value of Mach number is used to determine number density using eq.-\eqref{eq:n}. Jet is operated under different reservoir pressures to change the extent of rarefaction. The values of reservoir pressure and corresponding background pressure for continuous operation of jet are given in table \ref{table:exp_cond}. 

\begin{table}[h]
\begin{center}
\begin{tabular}{|c|c|c|c|c|}
\hline
		 	& $a$		 		& $b$		         & $c$		 		  & $d$		    	   \\ \hline
$P_0$ mbar 	& $100\pm1.5$ 		& $50\pm1$           & $10\pm0.1$ 		  & $5\pm0.01$  	   \\ \hline
$P_b$ mbar  & $1.25\pm0.1e^{-2}$& $6.14\pm0.1e^{-3}$ & $1.24\pm0.05e^{-3}$& $3.86\pm0.1e^{-4}$ \\ \hline  
$n_b/n_0$	& $1.26e^{-4}$ 	    & $1.22e^{-4}$ 	     & $1.24e^{-4}$ 	  & $7.70e^{-5}$       \\ \hline
\end{tabular}
\end{center}
\caption{\label{table:exp_cond} Background pressures $P_b$ for continuous jet at different reservoir pressures $P_0$; a) 100 mbar, b) 50 mbar, c) 10mbar, d) 5mbar.  Variations of values determines the precision of measurement. Additional uncertainty in measurements arise from the accuracy of the gauges used. For $P_0$ accuracy is $10\%$ and for $P_b$ it is $30\%$. Both precision and accuracy is accounted in estimating the density. $n_b/n_0$ is the density ratio of the expansion. It represent the minimum possible normalized jet density ($n/n_0$). i.e. $(n/n_0)>(n_b/n_0)$}  
\end{table}

\begin{figure}[ht]
\centering
\includegraphics[width=0.45\columnwidth, trim=0cm 0cm 0cm 0cm, clip=true,angle=0]{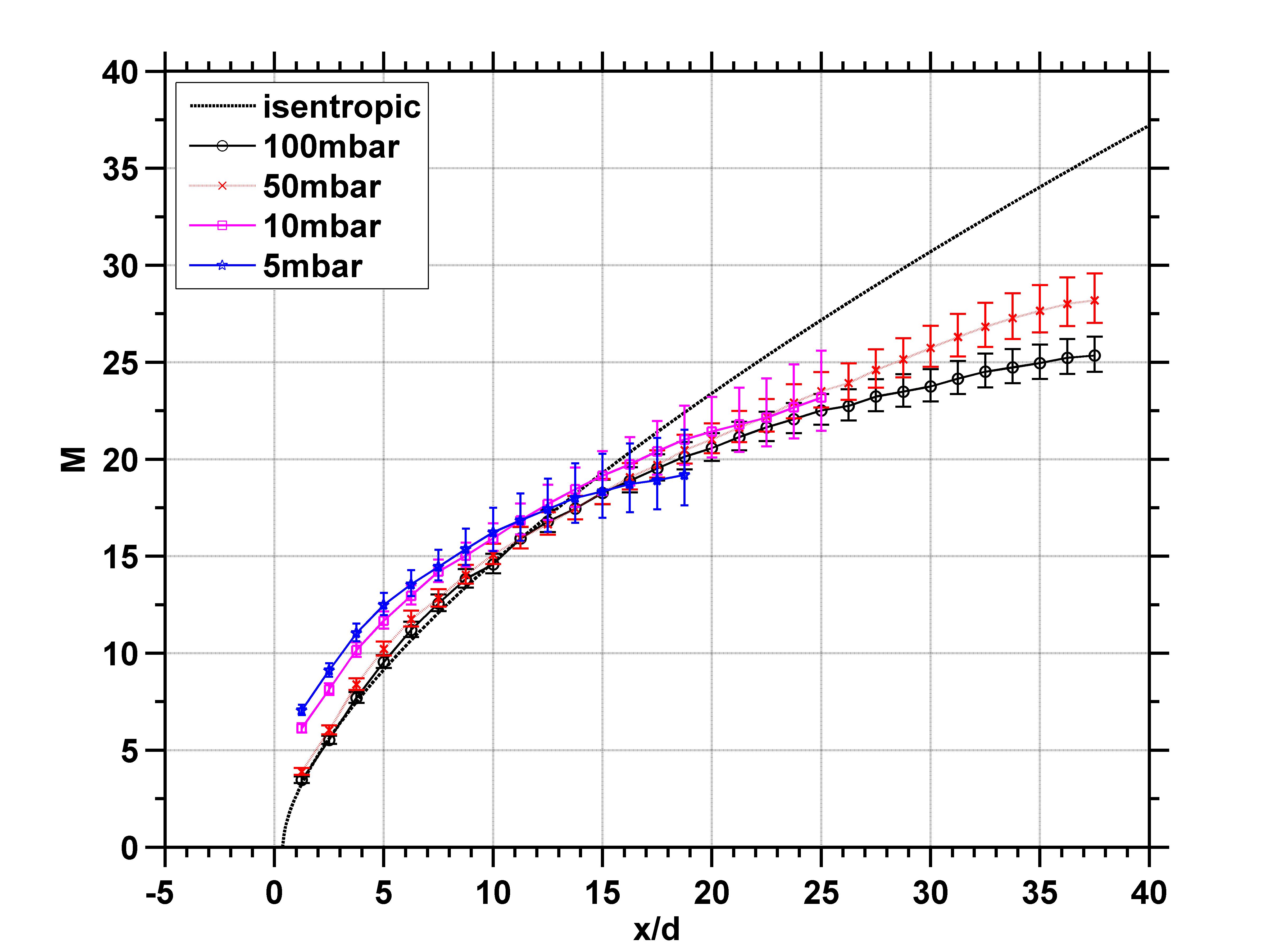}
\includegraphics[width=0.45\columnwidth, trim=0cm 0cm 0cm 0cm, clip=true,angle=0]{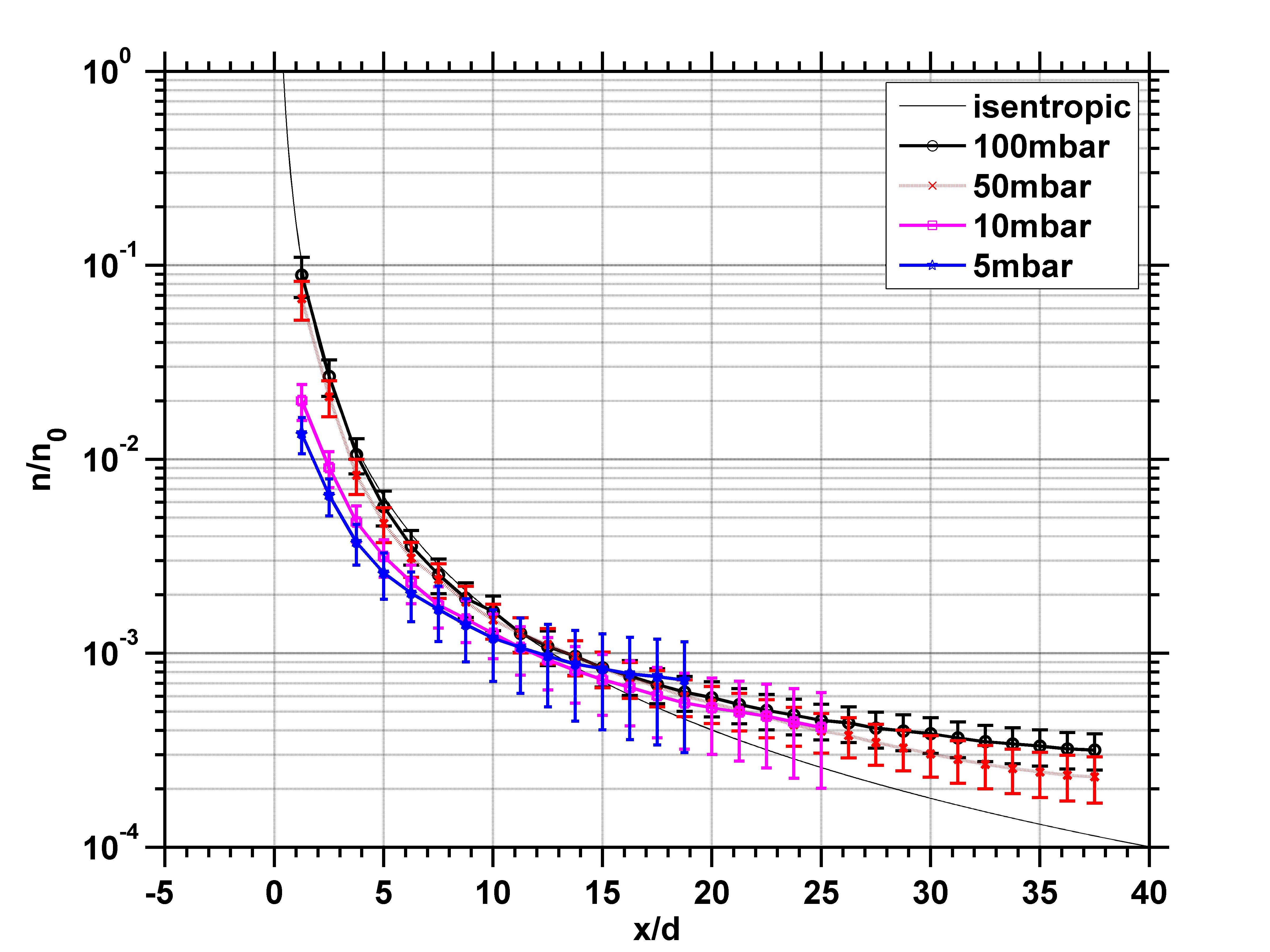}
\caption{\label{fig:exp_result_sonic} Axial Mach number(left) and number density(right)  calculated from impact pressure measured experimentally at different axial distances for different reservoir pressures. 
}
\end{figure}

\begin{figure}[ht]
\centering
\includegraphics[width=0.45\columnwidth, trim=0cm 0cm 0cm 0cm, clip=true,angle=0]{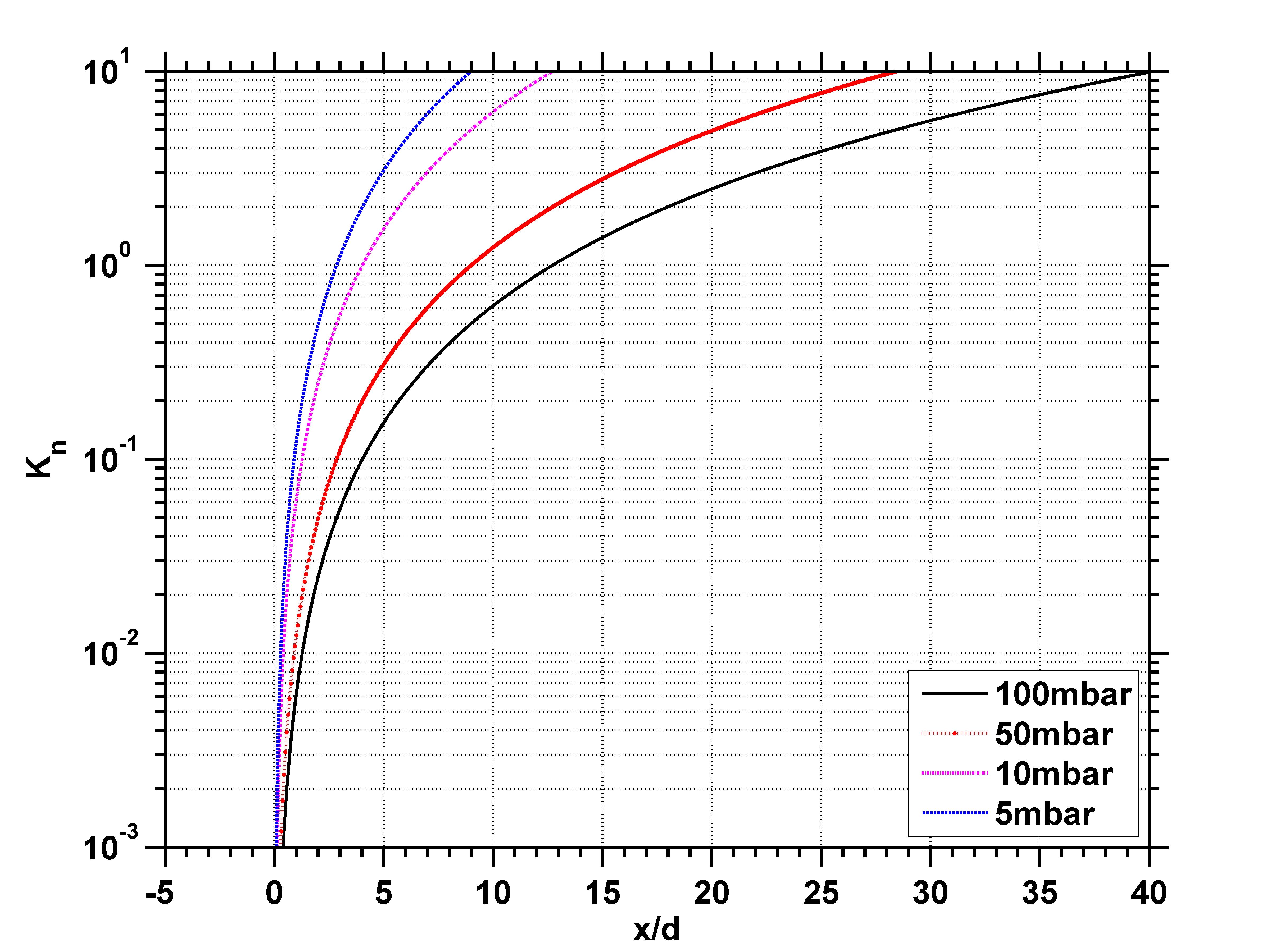}
\includegraphics[width=0.45\columnwidth, trim=0cm 0cm 0cm 0cm, clip=true,angle=0]{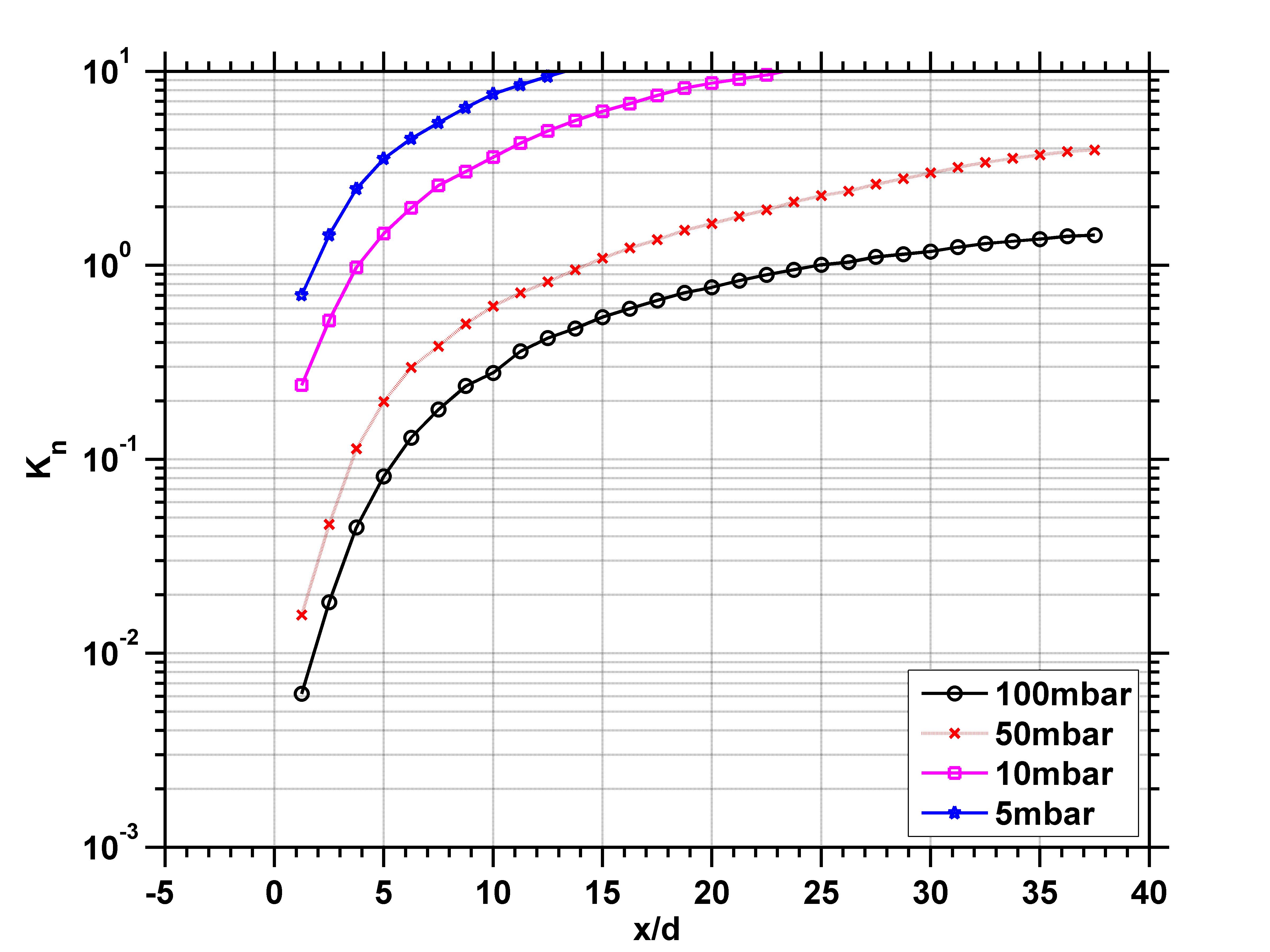}
\caption{\label{fig:exp_result_sonic_Kn} Theoretical values of Knudsen number inside the jet (left) and inside the pitot tube derived from the impact pressure(right).}
\end{figure}

Figure-\ref{fig:exp_result_sonic} shows the measured values of Mach number and density normalized to reservoir pressure for four different values of reservoir pressure given in the table \ref{table:exp_cond}. Viscous effects (in experiments) to calculate Mach numbers are taken into account by considering the reduced flow diameter ($0.894 d$) calculated from experimental flow rate ($0.80 \dot{N}_{theoritical}$) using eq.-\eqref{eq:flowrate}. The experiments are also carried out at $P_0$ 25 mbar and 1mbar, however, these values are not shown here for the sake of clarity. The error bars shown in the figure indicate cumulative errors in the measured values, accounting for precision in the measurements and the accuracy of gauges (connected to pitot tube $10\%$). A back-calculation of Mach numbers using RPT equation is iterative in nature, the upper and lower bounds of the errors are determined by using maximum and minimum limits of the measured values of impact pressure.
The deviations in the values of Mach number and/or number density can be explained using the flow conditions (Knudson number) in the pitot tube. As the density inside the pitot tube is has contribution both from dynamic and static parts of the flow, the density of the gas inside the pitot tube is always higher than the absolute value of the density in the jet. As a result, the collision frequency in the pitot tube is higher than that inside the jet. Hence, $K_{n(jet)}>K_{n(pitot)}$. Thus, pitot tube gives relatively accurate values even if the flow conditions in the jet slightly tend towards the rarefied regime. $K_{n(jet)}$ is calculated for the length scale of nozzle ($1 mm$) while $K_{n(pitot)}$ is calculated for the length scale of pitot tube ($0.5 mm$) and are shown in figure-\ref{fig:exp_result_sonic_Kn}. It can be seen that for all the values of $P_0$, flow at far expansion region is always molecular. Hence, the axial density gradually accommodates to the background density which is approximately four orders of magnitude lower ($\approx 2.5 \times 10^{-4} n_0$ ; table \ref{table:exp_cond}).

For jets at 50mbar and 100mbar, number density values agree with the theory for $x/d<12$, after which the effects of penetration of the background become apparent and a gradual increase is observed as the jet density accommodates to the background density. From figure-\ref{fig:exp_result_sonic_Kn} we see that this agreement is until $K_n$<0.3 (at $x/d \approx 12$). Along with penetration effects, measurement accuracy also deteriorates in the far axial distances ($x/d>15$) as the flow becomes rarefied and the pitot tube gives smaller than expected values because of the reduced throughput from the interference effect as discussed earlier. As a result, the density for 50 mbar jet is slightly smaller than 100 mbar jet (figure-\ref{fig:exp_result_sonic}) despite both having the same density ratios ($n_b/n_0$) for expansion (table-\ref{table:exp_cond}).

For jets at 5 mbar and 10 mbar, under-prediction of density for $x/d<12$ is again due to the rarefied nature of the jet. However, the trend shows over- prediction for $x/d>15$. For jets at 10 and 5 mbar, the measured value of impact pressure approaches the limit of measurement of the pitot tube assembly. As a result, pressure starts saturating at $5 \times 10^{-3}mbar$ which translates to minimum measurable number density ($n_{lim}$) to $1.25 \times 10 ^{20} molecules/m^3$. For 10 mbar jet, the normalized density corresponding to this value is $n_{lim}/n_{0(10mbar)} = 5 \times 10^{-4}$ and for 5mbar it is $n_{lim}/n_{0(5mbar)} = 1 \times 10^{-3}$. As a results, for jets at 10 mbar and 5 mbar the density values are over-predicted for $x/d>12$ and $x/d>16$, only because of the limits of the measurement system. Of course, the limit of measurement system (pitot tube assembly) is not due to measuring limit of the gauge but is imposed due to decrease in conductance of the pitot tube assembly when the flow inside the assembly becomes molecular. Thus, the gauge at the end of the assembly does not respond to the pressure variation at the tip of the tube.

Briefly, above discussion gives a general idea of behavior of pitot tube in transient and rarefied regime. Based on the level of rarefaction, the measurement error can be an order of magnitude. Additionally, deviations from ideal isentropic conditions are caused due to gradual accommodation of the rarefied jet to the background pressure. 
As a result when using the pitot tube to measure the jet density for a supersonic molecular nozzle operating in rarefied regime, it only makes sense to compare the measurements qualitatively. In addition at present there is no available model that accurately describes rarefied expansion through a supersonic nozzle. Hence, it is difficult to validate the experimental measurements in case of a supersonic nozzle. Hence trends in axial density are compared qualitatively with the results from DSMC simulation.

\subsection{Comparison of parabolic nozzle with conical counterparts}

\begin{figure}[ht]
\centering
\includegraphics[width=0.47\columnwidth, trim=0cm 0cm 0cm 0cm, clip=true,angle=0]{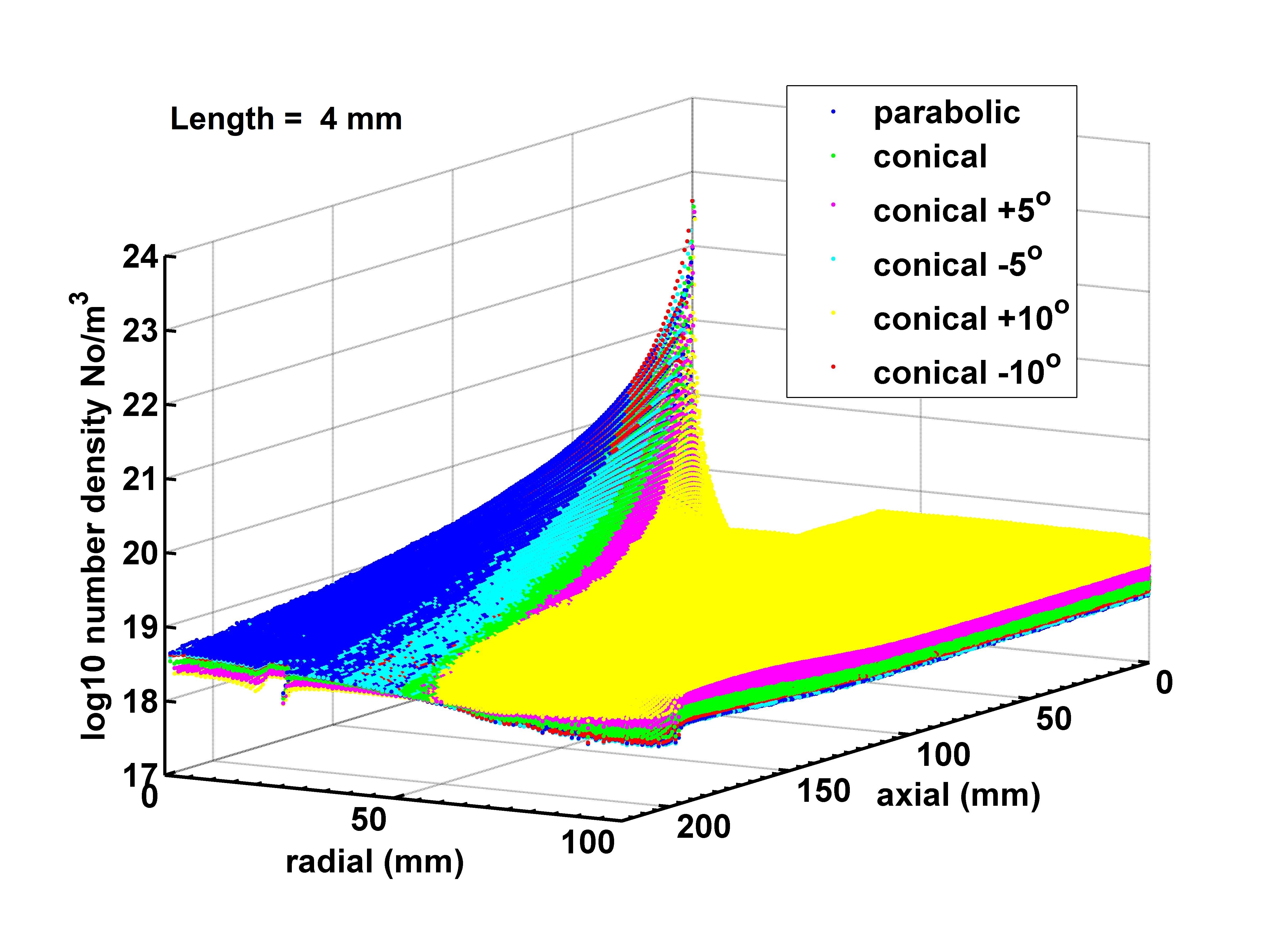}
\includegraphics[width=0.47\columnwidth, trim=0cm 0cm 0cm 0cm, clip=true,angle=0]{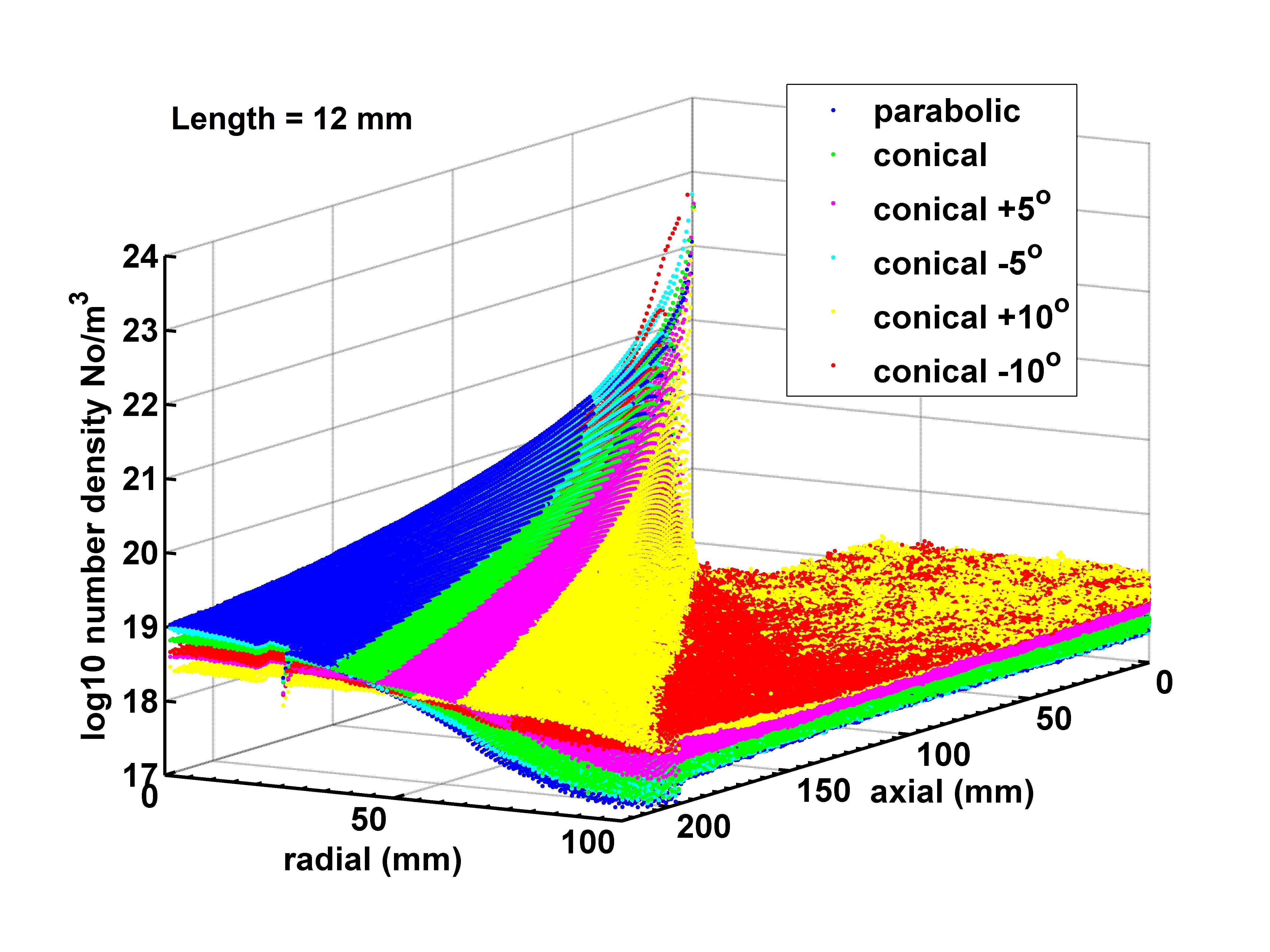}
\includegraphics[width=0.47\columnwidth, trim=0cm 0cm 0cm 0cm, clip=true,angle=0]{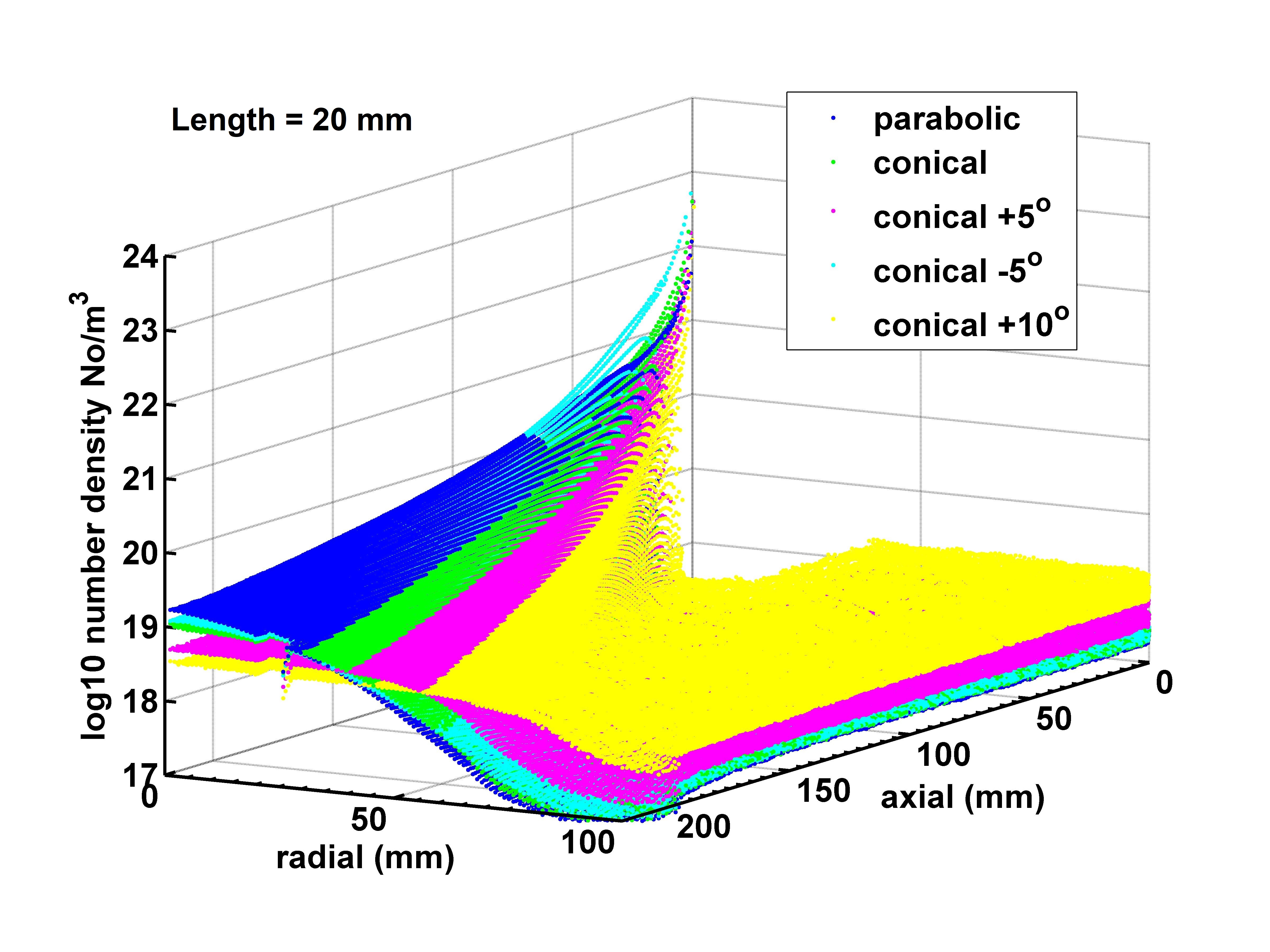}
\caption{\label{fig:comp} Comparison of axial density distribution for $l_e =4,12,20$ mm, $R_n=0.4$. Parabolic nozzle shows superior performance over conical nozzle. Additionally, equivalent conical nozzle can be the second best option to produce maximum axial density.}
\end{figure} 
 
Parabolic and conical nozzles used in the experiment are individually mounted on the pulse valve having orifice diameter of 1 mm. As was previously mentioned, all supersonic nozzles have 0.8 mm throat diameter. This is done to make sure that it is the nozzle orifice, rather than the pulse valve, where the supersonic expansion starts. The variation in throat diameters across different nozzles is of about 0.1 mm. As a result, the throughput of each nozzle differs. Hence, this may result in slight discrepancy in the absolute density. In order to guarantee that the expansion inside the nozzle is transient and molecular, the reservoir pressure ($P_0$) is maintained at 10 mbar throughout all measurements. However, due to measurement limitations as described earlier, experiments cannot provide reliable results for pressure below $P_0$<10 mbar. Simulations using the DSMC code are run for identical conditions as the experiment, as illustrated in section-\ref{sec:sim}

We studied the performance of nozzles using simulations for three different lengths of the expansion section ($l_e$), 4 mm, 12 mm, and 20 mm; however, experimental measurements were only performed for the 20 mm length. At each length, we compare the performance of the parabolic nozzle with that of equivalent conical nozzle. We would like to mention that from equivalent nozzle we mean a conical nozzle with the same angle subtended by the exit diameter at the throat as a parabolic nozzle. In addition, nozzles with same throat diameter and lengths but different half cone angles are also taken into account.

Figure-\ref{fig:comp} shows the number density trends obtained from DSMC simulations for all the nozzles.
For the same flow rate, the parabolic nozzle shows higher axial density in comparison to other conical nozzles of different lengths. The flat density regions arise from accumulation of background inside the flow domain due to the jet. Even though same effects are observed in the experiments, absolute values should not be compared as it is not possible to introduce accurate value of the experimental pumping rate in the simulations. However, these do not affect the results presented in this work as the region of interest is on the flow axis.  

\begin{figure}[ht]
\centering
\includegraphics[width=0.47\columnwidth, trim=0cm 0cm 0cm 0cm, clip=true,angle=0]{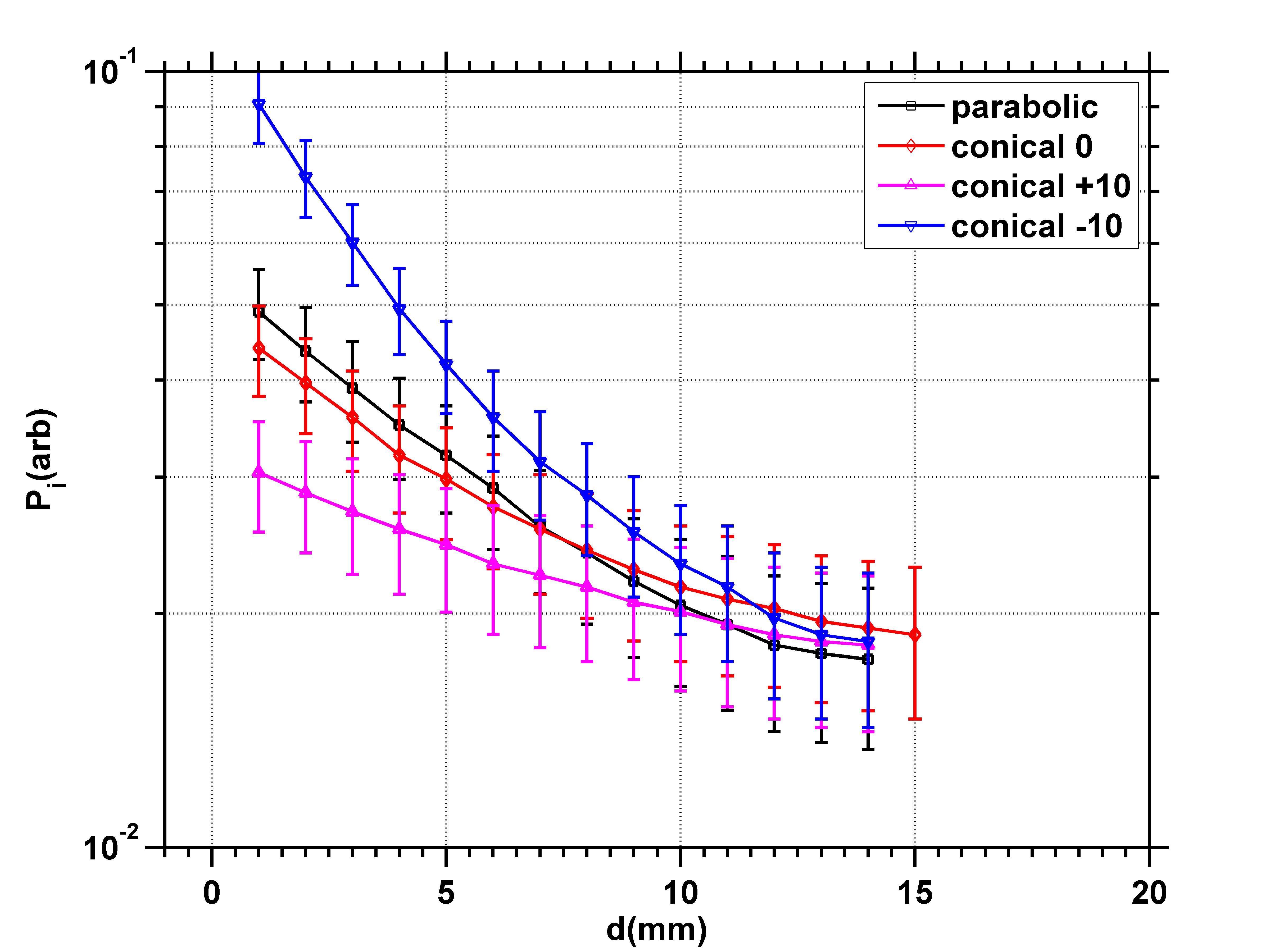}
\includegraphics[width=0.47\columnwidth, trim=0cm 0cm 0cm 0cm, clip=true,angle=0]{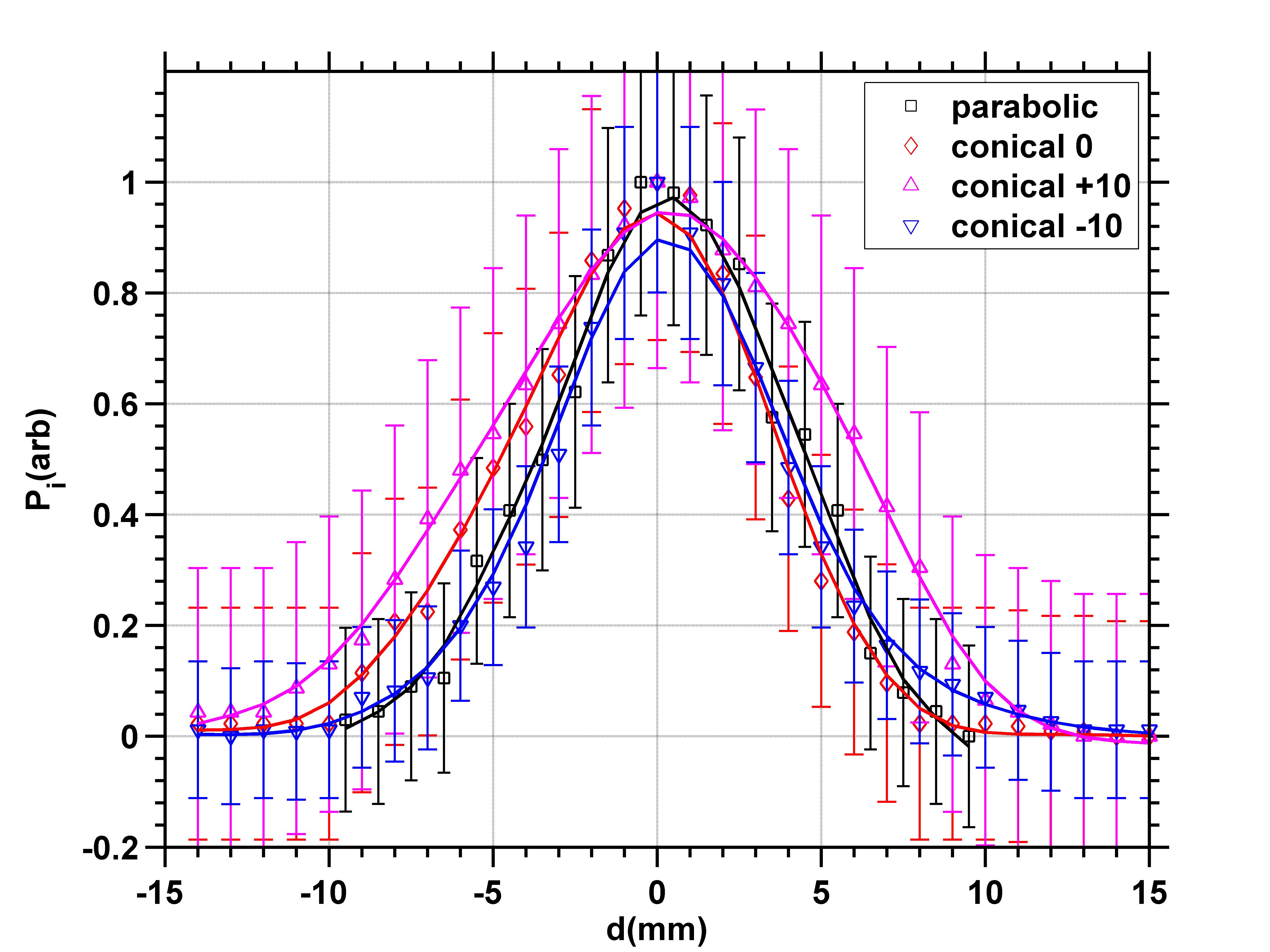}
\caption{\label{fig:exp_profiles} Axial density profiles (left) and radial spread of jet at 10 mm (right) for jet through different nozzle}
\end{figure} 

It is straight forward to check the optimum performing nozzle in simulation from absolute density. Since absolute density could not be accurately measured in experiments, we use a different method to ascertain nozzle performance.
Axial density following the nozzle exit for a molecular jet is influenced by the absolute density at the exit as well as the expansion angle of jet at the exit, because the path of the molecules is unchanged once they leave the nozzle. To produce an intense and narrow jet, the exit density should be high and expansion angle should be low. Exit diameter controls the exit density, and the axial density fall-rate may be used to estimate the expansion angle. The expansion angle of the jet will be greater and vice versa depending on the density fall-rate. In the experiment, we show the fall-rate of axial density using impact pressure since both follow the same pattern. We also determine the lateral profile of the jet at 10 mm from the nozzle. When we compare this with the exit diameter of the nozzle, we get the expansion angle of the jet. 

Figure-\ref{fig:exp_profiles} shows experimental measurements of impact pressure and lateral density carried out for one parabolic and three conical nozzles (equivalent $0^o$, wide $+5^o$ and narrow $-5^o$). 
Due to the detection limit of the pitot tube assembly, we are able to measure impact pressure reliably only upto a distance of 15 mm. Errors are estimated using the same approach as in case of the sonic nozzle. The inaccuracies are substantially higher at larger distances, making it impossible to draw any definite conclusion.
For d<15mm where errors are relatively small, the axial density fall for parabolic and equivalent conical nozzles is nearly identical. Given that both nozzles have same exit diameter, their exit densities are also the same. This demonstrates that both jets have comparable lateral expansion after leaving the nozzle. However, we can observe that the jet from the parabolic nozzle is a little bit narrower. As a result, albeit not significantly, parabolic nozzle provides somewhat higher directionality than a conical nozzle of equal size. 
In case of a narrow conical nozzle, axial density at exit is higher due to small size of the exit diameter. However, the axial density fall-rate is higher resulting in poor performance at longer distances. The lateral spread of the jet is also the same, if not slightly smaller than that of a parabolic nozzle. This is mainly due to the fact that the jet had a narrower profile at the nozzle exit than a parabolic nozzle. The fact that the jet losses its ability to maintain a narrow profile at the measuring distance indicates that the expansion angle is greater than that of the parabolic nozzle jet. As a result, the jet will have smaller axial density at extended distances. 
On the other hand, for wide conical nozzle, the jet has a relatively flat fall-rate. The exit density, however, is lower due to the wide exit diameter, and the lateral profile is notably broad. 
 
We see that the parabolic nozzle perform significantly better than wider and narrow conical nozzles but the improvement is rather small compared to its conical equivalent within the experimentally measured distances. Thus, a conical equivalent nozzle may also be employed as a geometrically simple alternative. However, the dimensions of the conical equivalent (length and exit diameter) are derived from the parabolic profile, demonstrating the significance of parabolic nozzle. Here we would like to point out that the comparison would have been more distinctive at longer distances from the nozzle, which, however, is not possible from the present experiment. Nonetheless the study demonstrates that the performance of additively manufactured parabolic nozzle is comparable or slightly better as compared to conical equivalent.

\section{Conclusion}\label{sec:con}

In this work, we describe a method for producing a narrow supersonic free jet with high axial density across extended axial lengths using sub-mm throat size using molecular nozzles. A parabolic profile based on the virtual source concept of free expansion and intended to work under transient and rarefied flow conditions is presented.
Nozzles used in the experiment are made using ABS by additive manufacturing method. Geometrical precision and vacuum performance is found to be satisfactory for room-temperature jets once interior surfaces are
smoothed properly.

DSMC simulations and experiments are used to compare the performance of the parabolic nozzle with a set of conical nozzles with different opening angles and lengths at the same flow rate. Density profiles for rarefied flow are measured experimentally using a pitot tube. The measurement uncertainties when employing a pitot tube for rarefied flows are assessed using a free jet through a sonic nozzle working in both the continuum and rarefied regimes. We observe that the pitot tube gives reliable results for $K_n<0.1$. However, for higher $K_n$, the measured density underestimates and can result in error of an order of magnitude at $K_n=1$

According to simulations, parabolic nozzle shows the highest centerline density for all nozzle lengths when compared to similar conical nozzles with varying opening angles. Experiment results demonstrate that parabolic nozzle performs slightly better than the equivalent conical nozzle. As rigorous conclusive comparisons could not be made due to the limitations of experiments, details in the nozzle performance need to be investigated in future. However, the present study shows positive trends regarding the parabolic nozzle to generate directional free jets with high axial density.

Briefly in the present work we report expansion of jet using conical and additively fabricated parabolic nozzle. Impact pressure is measured with the help of pitot tube. Their performance is compared from experiment as well from DSMC simulations and theoretical model. Additively manufactured parabolic  nozzle has performance comparable or slightly better when compared to conical nozzles. We believe that present study will be informative in optimizing the nozzles for center-line density for rarefied jets.
 
\section*{Data Availability Statement}

The data that support the observations of this study are available from the corresponding author upon reasonable request.

\section*{Author Declaration}

The authors have no conflict to disclose

\appendix
\section{Derivation of the flow rate}\label{appA}

For supersonic as well as sonic nozzles, choked flow condition occurs at the throat section where the cross-sectional area is minimum. Hence, the flow velocity at the throat is sonic (M=1). The mass flow rate through the nozzle can be determined form the mass flow rate at the throat.

\begin{equation}
\dot{m}=\rho^*A^*v^*=\frac{\rho^*}{\rho_0}\rho_0A^*v^*
\label{appeq:a1}
\end{equation}

From equations of isotropic expansion at the throat (M=1) 
\begin{equation}
\frac{\rho^*}{\rho_0}=\left(1+\frac{\gamma-1}{2}\right)^\frac{1}{1-\gamma},
\rho_0=n_0\frac{MW}{N_A}, 
v^*=\sqrt{\gamma R T^*}
\label{appeq:a2}
\end{equation}

where, $MW$ = molecular weight, $N_A$=Avagadro's constant $R$=characteristic gas constant

Using \ref{appeq:a1} and \ref{appeq:a2} 
\begin{equation}
\dot{N}=\dot{m}\frac{N_A}{MW}=\left(\frac{\gamma+1}{2}\right)^{\frac{1}{1-\gamma}} n_0 A^* \sqrt{\gamma R T^*}
\end{equation}

We can relate flow rate to reservoir condition and thermal velocity using $T_0/T^*=(\gamma+1)/2$ for $M=1$ and  $\alpha_0=\sqrt{2 R T_0}$ respectively. Thus we get

\begin{eqnarray}
\dot{N}=f(\gamma)n_0\alpha_0\pi (d/2)^2
\label{appeq:a4}
\\
where f(\gamma) is \sqrt{\frac{\gamma}{2}} \left(\frac{\gamma+1}{2}\right) ^{-\frac{1}{2}\frac{\gamma+1}{\gamma-1}}
\label{appeq:a5}
\end{eqnarray}

\section{measuring flow rate by pressure rise method}\label{appB}

To calculate the experimental flow rate, rise in the background pressure is measured for a known value of the gas injection period. For redundancy, measurements are performed for two different background pressures $1\times10^{-2}$ mbar and 1 mbar. Care has been take to ensure that pressure inside the measurement vessel remains unchanged for the duration larger than the gas injection period. Assuming ideal gas equation, the rise in pressure relates directly to the mass flow rate

\begin{equation}
\dot{m}=\frac{\Delta m}{\Delta t}=\frac{\Delta P V/RT}{\Delta t}
\end{equation}

Here, $\Delta t$ is gas injection time. Number flow rate ($\dot{N}$) can be determined by dividing mass flow rate by mass of each molecule ($MW/N_A$)

\begin{equation}
\dot{N}=\frac{\dot{m}N_A}{MW}
\end{equation}

Experimentally measured flow rate is approximately $80\pm 5\%$ of the theoretical prediction given by eq.-\ref{eq:flowrate}. Hence, correction to flow tube radius is applied to calculate Mach number using eq.-\ref{eq:mach} when used to compare with pitot tube measurements. Similarly when using eq.-\ref{eq:n_bjrnk}, correction is applied by using experimently measured value of $\dot{N}$.

\bibliography{references}

\end{document}